\documentclass[%
 reprint,
 amsmath,amssymb,
 aps,
 prb,
]{revtex4-2}
\usepackage{color}
\usepackage{graphicx}
\usepackage{dcolumn}
\usepackage{bm}
\usepackage{mhchem}
\usepackage{comment}

\begin{document}


\title{Physics of Phonons in Systems with Approximate Screw Symmetry}

\author{Hisayoshi Komiyama}
\author{Tiantian Zhang}
\author{Shuichi Murakami}
\affiliation{
  Department of Physics, Tokyo Institute of Technology, 2-12-1 Ookayama, Meguro-ku, Tokyo 152-8551, Japan
}

\date{\today}

\begin{abstract}

Properties of systems with exact $n$-fold screw symmetry $(n=2, 3, 4, 6)$ have been well studied because they can be understood in terms of space groups.
On the other hand, existence of materials with {approximate screw symmetries, such as 7-fold and 10-fold screw symmetries, }has been predicted.
In this paper, we study properties of phonons in crystals with approximate screw symmetries, {which will lead to unique and new physical phenomena}.
In a crystal with an approximate screw symmetry, we propose a method to extract information of pseudoangular momentum of phonons, which is a quantum number characterizing the 
properties of phonon modes under screw symmetry, {based on the fact that the information of the quantum numbers defined under exact screw symmetry partially remains in the eigenvectors of approximate screw symmetric systems.}
As a preparation, we study a one-dimensional crystal with partially broken translation symmetry to have an enlarged unit cell, and we show how to extract information of a quantum number corresponding to the pseudoangular momentum, by studying a relative phase between neighboring atoms.
We also extend this method to systems with an approximate screw symmetry, and discuss properties of the pseudoangular momentum.
We apply this method to results of our first-principle calculations  {on} candidate materials with an approximate translational symmetry  {or} with an approximate screw symmetry, and show how this approximate symmetry is reflected in the phonon wavefunctions.

\end{abstract}

\maketitle

\section{Introduction}

A phonon is a quasiparticle of quantized lattice vibrations and has long been studied to account for various physical properties such as specific heat, transport and electric resistance.
Recently, chiral phonons, in which atomic motions have rotational components with chirality, have been predicted theoretically \cite{zhang2015chiral} and observed experimentally \cite{zhu2018observation}.
In addition, chiral phonons are expected to have a variety of applications and have been studied extensively \cite{liu2017pseudospins, chen2018chiral, gao2018nondegenerate, xu2018topological, xu2018nondegenerate, chen2019chiral, li2019emerging, li2019momentum, liu2019valley, chen2019entanglement, delhomme2020flipping,zhang2020chiral, grissonnanche2020chiral, chen2021probing, suri2021chiral, yin2021chiral, maity2022chiral,zhang-chiralHall}.
Furthermore, the concept of chiral phonons has been studied in crystals with screw symmetry, and interactions with other particles with chirality and experimental methods have been proposed \cite{zhang2018double, kishine2020chirality, ishito2021truly, chen2021chiral, zhang2021chiral, chen2021propagating, li2021observation, juneja2021quasiparticle, wang2022chiral, skorka2022chiral, chen2022chiral, choi2022chiral,PhysRevB.98.241405,
	PhysRevB.80.125407,
	PhysRevB.85.035436,
	PhysRevMaterials.5.085002,
	PhysRevB.82.161402,
	JUNEJA2021100548,
	PhysRevB.61.3078}.

So far, properties of chiral phonons have been studied in materials with exact rotational and screw symmetries.
In this case, the chiral phonons are characteristic by a quantity called pseudoangular momentum \cite{bovzovic1984possible, zhang2015chiral, zhang2021chiral}, which is used to discuss interactions with other particles.
The definition of the pseudoangular momentum is based on rotation or screw symmetry.
As crystallographic symmetries, {those} symmetries are limited to $n$-fold rotation and screw {rotation} symmetries with $n=2, 3, 4, 6$.
On the other hand, materials with approximate screw {symmetries} such as $7_2$ screw and $10_7$ screw symmetries are predicted.
Since pseudoangular momentum cannot be defined for {the} approximate screw symmetry, {pseudoangular momentum cannot be defined directly in such systems}. However, we can expect that materials with approximate symmetries have the same {pseudoangular} momentum information as those with strict screw {symmetries}.
Therefore, in this paper, we discuss how to extract information of pseudoangular momentum for the phonons in systems with approximate symmetry, starting with approximate translation symmetry, and then extending to approximate screw rotation symmetries. Furthermore, we also discuss characteristic behaviors of the phonon eigenmodes in these materials with approximate symmetries.
{Such properties under approximate symmetry affect physical processes involving various particles/quasiparticles, such as electronic processes involving multi-phonons/photons and exciton scattering processes. In systems with exact symmetry, symmetry restricts such processes as selection rules, and even when the symmetry is not exact but approximate, selection rules remain valid. While exact screw rotation symmetries are limited in crystals to twofold, threefold, fourfold and sixfold ones due to the restriction of translational symmetry, an approximate screw symmetry leads to a richer variety of multifold screw symmetries such as 7- and 10-fold ones, and they will lead to selection rules that are absent for exact screw symmetries. Furthermore, anisotropy in transport properties involving chiral phonons will also be reflected the approximate screw symmetry. Thus, the approximate screw symmetries affect phonon-related transport phenomena.}

This paper is organized as follows.
In Sec.~\ref{sec:1D chain}, we study phonons in a one-dimensional crystal with {approximate translation symmetry, i.e., slightly broken translation symmetry to have an enlarged unit cell. From the relative phases between neighboring atoms in the case with the exact translation symmetry, we show how to obtain the quantum number, which correponds to the pseudoangular momentum in 3D systems, for crystals with approximate translation symmetry
In Sec.~\ref{sec:3D helix}, we show how to extract the phonon pseudoangular momenta for crystals with approximate screw symmetries, which is also obtained from the exact screw symmetries, and discuss their properties. }
In Sec.~\ref{sec:first-principles calculation}, we apply this method to first-principle calculations on phonon modes in candidate materials and show that the approximate symmetries are well reflected in phonon wavefunctions.
In Sec.~\ref{sec:conclusion}, we summarize this paper. 

\section{\label{sec:1D chain} One-dimensional phonon}

\begin{figure}
  \includegraphics[clip,width=\linewidth]{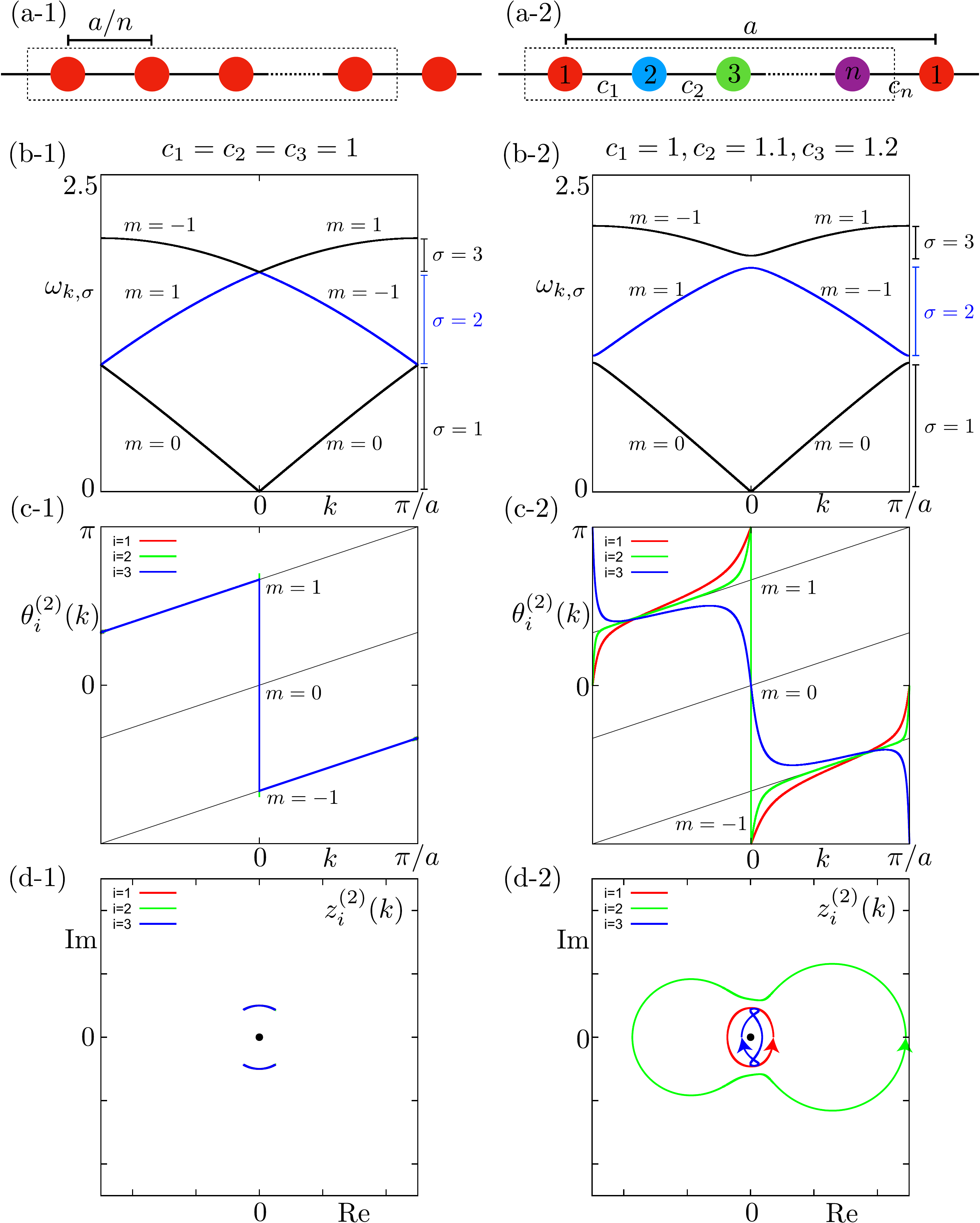}
  \caption{\label{fig:phase}
  Phase differences between neighboring atoms for the one-dimensional phonon system.
  (a) Schematic picture of the system (a-1) with and (a-2) without $\hat{T}_{a/n}$ translation symmetry.
  (b-1), (b-2) Band dispersions $\omega_k$ for the systems with (b-1) exact translational symmetry $\hat{T}_{a/3}$ ($c_1=c_2=c_3=1$) and (b-2) broken translational system $\hat{T}_{a/3}$ ($c_1=1$, $c_2=1.1$, $c_3=1.2$).
  (c-1), (c-2) Relative phase $\theta_{i}^{(\sigma=2)}(k)$ of the second band from the bottom and
  (d-1), (d-2) Trajectory of the complex number $z_{i} (k)$ of the second band from the bottom ($\sigma=2$).
  (c-1), (d-1) correspond to the case with exact translational symmetry in (b-1), and (c-2), (d-2) to the case with broken translation symmetry $\hat{T}_{a/3}$ in (b-2).
  }
\end{figure}
In this section, we study {phonon quantum number} in one-dimensional crystals with partially broken translational symmetry, {i.e., approximate translation symmetry,} such that the unit cell is enlarged by a factor $n$ ($n$ : integer), and discuss phase properties of the phonon eigenvectors.
We consider a one-dimensional chain of atoms with equal spacing $a/n$ ($a$: constant, $n$: integer), and the atoms can oscillate only along the chain direction. We begin with the case with all the atoms being equivalent, leading to a translation symmetry by $a/n$ (Fig.~\ref{fig:phase} (a-1)).

Then we perturb the system while preserving a translation symmetry by $a$ (Fig.~\ref{fig:phase} (a-2)).
Thus, the size of the resulting unit cell becomes $a$, having $n$ atoms.
The Hamiltonian of the one-dimensional phonon system is 
\begin{align}
  \label{eq:1D chain H}
  H = \sum_{l, l'} \left( \frac{1}{2} \bm{p}_{l}^{\mathrm{T}} \bm{p}_{l} \delta_{l.l'} + \frac{1}{2} \bm{u}_{l}^{\mathrm{T}} K_{l,l'} \bm{u}_{l'}  \right) ,
\end{align}
where
$\bm{u}_{l} = \left( u_{l,1}, u_{l,2}, \cdots, u_{l,n} \right)^{\mathrm{T}}$, 
$u_{l,i}$ is a displacement of the $i$th atom in the $l$th unit cell multiplied by the square root of the mass of the atom, 
$p_{l,i}$ is a conjugate momentum corresponding to $u_{l,i}$. The atoms are mutually connected by springs, and
$K_{l, l'}$ is a mass-weighted force constant matrix.

Because of the translation symmetry by $a$, the Hamiltonian $H$ commutes with a translation operator $\hat{T}_{a}$ by length $a$: $\left[ H, \hat{T}_{a} \right] = 0$.
Therefore, from the Bloch's theorem, the eigenfunction $\psi(x)$ of the Hamiltonian $H$ satisfies $\hat{T}_{a} \psi(x) = e^{-i k a} \psi(x)$ ($\frac{-\pi}{a} < k \le \frac{\pi}{a}$), where $k$ is the Bloch wavenumber corresponding to the translation $\hat{T}_{a}$.
Furthermore, if the system preserves the translation symmetry by length $a/n$ represented by $\hat{T}_{a/n}$, $\left[ H, \hat{T}_{a/n} \right] = 0$ holds and the eigenfunction $\psi(x)$ satisfies $\hat{T}_{a/n} \psi(x) = e^{-i k' a/n} \psi(x)$ $\left( -\frac{n \pi}{a} < k' \le \frac{n \pi}{a} \right)$ where $k'$ is the Bloch wavenumber corresponding to the translation $\hat{T}_{a/n}$.
By using $\left( \hat{T}_{a/n} \right)^{n} = \hat{T}_{a}$, we can relate $k$ and $k'$ as $e^{i k a} = e^{i k' a}$, i.e $k' = k + \frac{2 m \pi}{a}$, ($m$: integer).
Therefore, we have
\begin{align}
  T_{a/n} \psi(x) = e^{-i (k a + 2 \pi m)/n} \psi(x) \ \left( \frac{-\pi}{a} < k \le \frac{\pi}{a} \right)
\end{align}
where $m$ is given by 
\begin{align}
  m \equiv \left\{\begin{array}{ll}
    0, \pm 1, \cdots, \pm (n-1)/2 & \text{for} \ n \ \text{odd} \\
    0, \pm 1, \cdots, \pm (n-2)/2, n/2 & \text{for} \ n \ \text{even}
  \end{array}\right. \pmod n .
\end{align}
This integer $m$ is regarded as a quantum number having information of the chain with exact $\hat{T}_{a/n}$ symmetry and is defined in terms of modulo $n$, as we discuss in the following. 
From now on, in this section, we discuss how the information of the quantum number $m$ remains in the eigenfunction when the translational symmetry $\hat{T}_{a/n}$ is slightly broken while keeping the translational symmetry $\hat{T}_{a}$.

We label the atoms in the unit cell of the one-dimensional system to be $1, 2, \cdots, n$, and we assume that the springs exist only between neighboring atoms (Fig.~\ref{fig:phase} (a-2)).
Then, the phonon dynamical matrix $D(k)$ for the system is written as 
\begin{align}
  D(k) = \begin{pmatrix}
    c_1 + c_n & -c_1 &  \cdots & 0 & -c_n e^{-ika} \\
    -c_1 & c_1 + c_2 &  \cdots & 0 & 0 \\
    \vdots & \vdots & \ddots & \vdots & \vdots \\
    0 & 0 & \cdots & c_{n-2} + c_{n-1} & -c_{n-1} \\
    -c_n e^{ika} & 0 & \cdots & -c_{n-1} & c_{n-1} + c_{n}
  \end{pmatrix} ,
\end{align}
where $k$ is the Bloch wavenumber corresponding to the translation $\hat{T}_{a}$ and has a value in range $-\frac{\pi}{a} < k \le \frac{\pi}{a}$.
Let $c_i$ $(i=1, \cdots, n-1)$ denote the spring constant between atom $i$ and atom $i+1$ and  $c_n$ denote the spring constant between atom $n$ and atom $1$ of the neighboring unit cell.
The crystal has the translational symmetry $T_{a/n}$ when the spring constants $c_i$ $(i=1, \cdots, n)$ are all equal, in which case the eigenvector of the dynamical matrix $D(k)$ for the $\sigma$-th eigenmode is 
\begin{align}
  \label{eq:exact eigenvector}
  \bm{\epsilon}_{k}^{(\sigma)} = \begin{pmatrix}
    1 \\
    e^{i (ka + 2m\pi)/n} \\
    \vdots \\
    e^{i (n-1)(ka + 2m\pi)/n}
  \end{pmatrix} .
\end{align}
Here the $i$-th component of the eigenvector $\bm{\epsilon}_{k}^{(\sigma)}$ of the dynamical matrix $D(k)$ represents the displacement of atom $i$.
Therefore, to exact the information of the quantum number $m$ from this eigenvector for the mode $\sigma$, we define a relative phase $\theta_{i}^{(\sigma)}(k)$ between atom $i$ and atom $i+1$ as
\begin{align}
z_{i}^{(\sigma)}(k) &= \left\{ \begin{array}{ll}
  \epsilon_{k, i+1}^{(\sigma)} / \epsilon_{k, i}^{(\sigma)} & i = 1, \cdots, n-1 \\
  \epsilon_{k, 1}^{(\sigma)} e^{ika} / \epsilon_{k, n}^{(\sigma)} & i = n
\end{array} \right. , \\
\theta_{i}^{(\sigma)}(k) &= \arg{z_{i}^{(\sigma)}(k)},
\end{align}
where we take the branch $-\pi < \theta_{i}^{(\sigma)}(k) \le \pi$.
When the system has the exact translational symmetry $\hat{T}_{a/n}$, from the eigenvector Eq~(\ref{eq:exact eigenvector}),
the values of $z_{i}^{(\sigma)}(k)$ and $\theta_{i}^{(\sigma)}(k)$ are
\begin{align}
  \label{eq:exact z}
  z_{i}^{(\sigma)}(k) &= e^{i \frac{ka + 2m\pi}{n}}, \\
  \label{eq:exact theta}
  \theta_{i}^{(\sigma)}(k) &= \frac{ka + 2m\pi}{n}
\end{align}
for $i=1, \cdots, n$, and the relative phase is a linear function with an intecept equal to 
$2 m \pi / n$.
On the other hand, when the translational symmetry $\hat{T}_{a/n}$ is broken by slightly changing the value of the spring constant $c_{i}$, we can also define the relative phases for the eigenvector.
Therefore, we can calculate the relative phases for the case with broken symmetry and extract the information of the quantum number $m$.

As an example, we calculate the relative phase $\theta_{i}^{(\sigma)}(k)$ in a specific model when we break the translational symmetry $T_{a/n}$ while keeping the translational symmetry $T_{a}$, {corresponding to the change from Fig.~\ref{fig:phase} (a-1) to (a-2).}
We take the number of atoms in the unit cell to be $n=3$.
In this calculation, the spring constants are varied from $c_1=c_2=c_3=1$ to $c_1=1, c_2=1.1, c_3=1.2$, in order to break the translational symmetry $T_{a/3}$ while keeping the translational symmetry $T_{a}$.
Then, the phonon dispersion changes from Fig.~\ref{fig:phase} (b-1) to Fig.~\ref{fig:phase} (b-2) and the relative phase $\theta_{i}^{(\sigma)}(k)$ of the $\sigma = 2$ band (the blue band in Figs.~\ref{fig:phase} (b-1)(b-2)) changes from Fig.~\ref{fig:phase} (c-1) to (c-2), and $z_i^{(\sigma)}$ changes
from Fig.~\ref{fig:phase} (d-1) to (d-2). 
From the results of Fig.~\ref{fig:phase}, in the case with the translational symmetry $T_{a/3}$, 
the relative phases $\theta_{i}^{(\sigma)} (i=1, 2, 3)$ are independent of $i$ and linear in $k$ and the complex number $z_{i}^{(\sigma)} (k)$ is on the unit circle in the complex plane (see Eq.~(\ref{eq:exact z})).
Next, when the translational symmetry $\hat{T}_{a/3}$ is slightly broken, the relative phase $\theta_{i}^{(\sigma)}(k)$ and the complex number $z_{i}^{(\sigma)}(k)$ remain almost the same with the case with strict translational symmetry $\hat{T}_{a/3}$.
Therefore, by calculating the relative phase for the case of slight symmetry breaking and comparing it with that for the case of strict symmetry, we can extract the information of the quantum number $m$ for each mode.
We show the results of the values of $m$ in Fig.~\ref{fig:phase} (b-1) and (b-2).

\begin{figure}
  \includegraphics[clip,width=\linewidth]{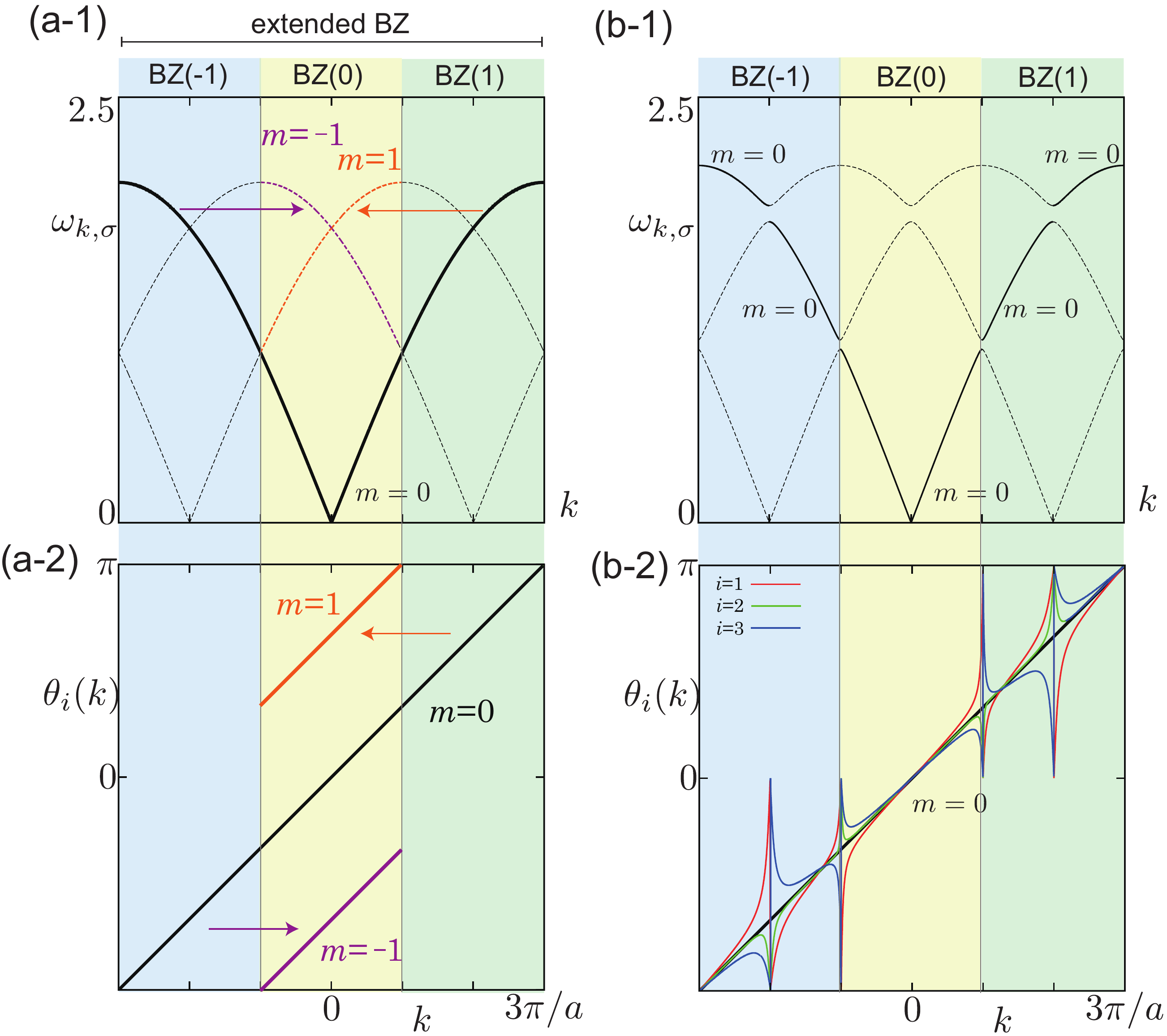}
  \caption{\label{fig:extend phase} 
  Phonon dispersions and relative phases between neighboring atoms in the extended Brillouin zone.
 (a-1) Band dispersion $\omega_{k, \sigma}$ and (a-2) Relative phase $\theta_{i}(k)$ for the system with translation symmetry $\hat{T}_{a/3}$ ($c_1=c_2=c_3=1$). 
 In the extended Brillouin zone (BZ),
 there is only one band with the quantum number $m=0$. 
 By folding the BZ into the region $-\pi/a\leq k \pi/a$ (light yellow region), the wavenumber $k$ is shifted and the quantum number $m$ changes accordingly. 
 The wavenumber $k$ in the region "BZ(1)" (light green region) is shifted by $-2\pi$, and the quantum number $m$ changes by $+1$. Similarly, 
  the wavenumber $k$ in the region "BZ(-1)" (light blue region) is shifted by $2\pi$, and the quantum number $m$ changes by $-1$. 
  (b-1)  
  Band dispersion $\omega_{k, \sigma}$ and (b-2) Relative phase $\theta_{i}(k)$ for the system without translation symmetry $\hat{T}_{a/3}$ ($c_1=1, c_2=1.1, c_3=1.2$).
  The relative phase holds the linear tendency before breaking the symmetry $\hat{T}_{a/3}$, but changes significantly at the time-reversal invariant momenta $k = l \pi / a$ ($l$: integer).
}
\end{figure}
We can naturally understand the behaviors of the quantum number $m$ using the extended Brillouin zone (BZ).
When the system has exact translational symmetry of $a/3$, the Brillouin zone is extended to $-3\pi/a\leq k \leq 3\pi/a$.
The band dispersion in the extended Brillouin zone for the systems with exact translation symmetry $\hat{T}_{a/3}$ ($c_1 = c_2 = c_3 = 1$) is shown in Fig.~\ref{fig:extend phase} (a-1).
In this zone scheme, there is only the phonon branch (solid line in Fig.~\ref{fig:extend phase} (a-1)), and 
its relative phase is given by 
\begin{align}
&\theta_i(k)=\frac{ka}{n},
\label{eq:kan}
\end{align}
which corresponds to $m=0$ in Eq.~(\ref{eq:exact theta}). It is seen from our numerical result for the relative phase shown as a black line in Fig.~\ref{fig:extend phase} (a-2).
When the unit cell size is enlarged to $a$, the Brillouin zone becomes $-\pi/a\leq k \leq \pi/a$ and the phonon bands are folded into this new Brillouin zone. Copies of the original phonon band appear as shown in broken lines in Fig.~\ref{fig:extend phase} (a-1). In this case, the right-neighboring Brillouin zone (BZ(1)) moves to the center Brillouin zone (BZ(0)) which adds $+1$ to the quantum number $m$, and the left-neighboring Brillouin zone (BZ(-1)) moves to BZ(0), adding $-1$ to the quantum number $m$. As a result, three branches appear within the new BZ.
From this argument, it shows that when $k$ increases and goes across the boundary of the Brillouin zone from $k=\pi/a$ to $k=-\pi/a$, the quantum number $m \pmod{n}$ increases by $+1$, {which is consistent with the discussions in Ref.~\cite{zhang2021chiral}.}
When the $\hat{T}_{a/3}$ symmetry is broken, the crossings in Fig.~\ref{fig:extend phase} (a-1) will become anticrossings as shown in Fig.~\ref{fig:extend phase} (b-1) for $c_1=1, c_2=1.1, c_3=1.2$. 
Within the extended zone scheme, the relative phases for the system with slightly broken 
translation symmetry $\hat{T}_{a/3}$ behave similarly to Eq.~(\ref{eq:kan}), as shown in Fig.~\ref{fig:extend phase} (b-2) expect for $k\sim 0$ and $k\sim\pi$. From this we can interpret that the band has quantum numbers like Fig.~\ref{fig:phase} (b-1) and (b-2).

On the other hand, near the time-reversal invariant momenta $k=0, \pi / a$, the relative phase $\theta_{i}^{(\sigma)} (k)$ and the complex number $z_{i}^{(\sigma)} (k)$ largely deviate from those in the system with $T_{a/n}$ symmetry.
In particular, at $k=0, \pi / a$, the relative phase $\theta_{i}^{(\sigma)}(k)$ quantizes to $0$ or $\pi$ because of the time-reversal symmetry.
To quantify this behavior, we define three quantum numbers from the relative phase $\theta_i^{(\sigma)}(k)$: 
\begin{align}
  p_{0, i}^{(\sigma)} &\equiv \theta_{i}^{(\sigma)}(0) / \pi \ \pmod{2}, \\
  p_{\pi, i}^{(\sigma)} &\equiv \theta_{i}^{(\sigma)}(\pi/a) / \pi \ \pmod{2}, \\
  w_{i}^{(\sigma)} & = \frac{1}{2\pi} \int_{k=-\pi/a}^{k=\pi/a} \frac{d \theta_{i}^{(\sigma)}}{d k} d k .
\end{align}
Here, $p_{0,i}^{(\sigma)}, p_{\pi,i}^{(\sigma)}$ take the values $0, 1$ modulo $2$ because $\theta_{i}^{(\sigma)}(0)$ and $\theta_{i}^{(\sigma)}(\pi/a)$ quantize to $0$ or $\pi$ due to the time-reversal symmetry.
On the other hand, $w_i^{(\sigma)}$ represents the number of times the curve $z_{i}^{(\sigma)}(k)$ $\left( -\pi/a < k \le \pi/a \right)$ travels counterclockwise around the origin in the complex plane, and is an integer.
The results of the calculation of these quantum numbers in the model of Fig.~\ref{fig:phase} are shown in Table~\ref{tab:phase}.
It is straightforward to show
\begin{align}
&w_i^{(\sigma)}\equiv   p_{\pi, i}^{(\sigma)}-  p_{0, i}^{(\sigma)} \pmod{2},
\label{eq:wpp}
\end{align}
because $\theta_i^{(\sigma)}$ is an odd function of $k$ (i.e. $\theta_i^{(\sigma)}(-k)=-\theta_i^{(\sigma)}(k)$), 
and indeed it holds in our results in Table \ref{tab:phase}.
\begin{table}
  \caption{\label{tab:phase}
  Quantum numbers $p_{0, i}^{(\sigma)}$, $p_{\pi, i}^{(\sigma)}$ and $w_{i}^{(\sigma)}$ of the model of Fig.~\ref{fig:phase}.
  }
  \begin{ruledtabular}
  \begin{tabular}{c|ccc|ccc|ccc}
    $\sigma$ &
    $p_{0, 1}^{(\sigma)}$ & $p_{0, 2}^{(\sigma)}$ & $p_{0, 3}^{(\sigma)}$ &
    $p_{\pi, 1}^{(\sigma)}$ & $p_{\pi, 2}^{(\sigma)}$ & $p_{\pi, 3}^{(\sigma)}$ &
    $w_{1}^{(\sigma)}$ & $w_{2}^{(\sigma)}$ & $w_{3}^{(\sigma)}$ \\
    \colrule
    1 & 0 & 0 & 0 & 1 & 0 & 0 & 1 & 0 & 0 \\
    2 & 1 & 1 & 0 & 0 & 0 & 1 & 1 & 1 & -1 \\
    3 & 0 & 1 & 1 & 1 & 1 & 1 & 1 & 0 & 0 \\
  \end{tabular}
  \end{ruledtabular}
\end{table}

We discuss properties of $p_{0,i}^{(\sigma)}$ and $p_{\pi,i}^{(\sigma)}$ with slightly broken translational symmetry $T_{a/n}$ in the following.
First, the eigenvector $\bm{\epsilon}_{k}^{(\sigma)}$ with the translational symmetry $T_{a/n}$ is specified by the quantum number $m$ and is given by
\begin{align}
  \bm{\epsilon}_{k}^{(\sigma)} = \left(\begin{array}{c}
    1 \\
    e^{i(ka+2m\pi)/n} \\
    \vdots \\
    e^{i(n-1)(ka+2m\pi)/n}
  \end{array}\right).
\end{align}
Meanwhile when $T_{a/n}$ is slightly broken, the eigenvectors should be real at time-reversal invariant momenta $(k = 0, \pi/a)$.
Therefore, a complex eigenvectors should be degenerate with its complex conjugate.
Let $m_1$ and $m_2$ be the quantum numbers for the two bands degenerate at the time-reversal invariant momentum $\left( k = 0,\ \pi / a \right)$; 
$\bm{\epsilon}_{k, m_1} = \bm{\epsilon}_{k, m_2}^{*}$.
As seen in Fig.~\ref{fig:extend phase} (a-1), this degeneracy exists in the systems with exact $T_{a/n}$ translation symmetry, and is due to the zone folding from the larger Brillouin zone ($-n\pi/a\leq  k \leq n\pi/a$), and it is lifted by slightly breaking the translational symmetry $T_{a/n}$.
Let $\bm{\epsilon}_{1}$ and $\bm{\epsilon}_{2}$ be the eigenvectors at the time-reversal invariant momentum when the translational symmetry $T_{a/n}$ is broken.
Within the zeroth order perturbation, they are linear combinations of $\bm{\epsilon}_{k, m_1}$ and $\bm{\epsilon}_{k, m_2} (=\bm{\epsilon}_{k, m_1}^{*})$.
Since the eigenvectors $\bm{\epsilon}_{1}$ , $\bm{\epsilon}_{2}$ are real vectors due to time-reversal symmetry, they are given by
\begin{align}
  \bm{\epsilon}_{1} = \left(\begin{array}{c}
    \cos{(\phi)} \\
    \cos{(\phi+\xi)} \\
    \vdots \\
    \cos{\left(\phi+(n-1)\xi\right)}
  \end{array}\right), \ 
  \bm{\epsilon}_{2} = \left(\begin{array}{c}
    \sin{(\phi)} \\
    \sin{(\phi+\xi)} \\
    \vdots \\
    \sin{\left(\phi+(n-1)\xi\right)}
  \end{array}\right), 
\end{align}
within the zeroth order perturbation, where $\phi$ is a constant determined from the spring constant matrix $K_{l,l'}$ and $\xi = (m_1-m_2) \pi / n$.
Therefore, for both the eigenvectors $\bm{\epsilon}_{1}$ and $\bm{\epsilon}_{2}$ at $k (= 0, \pi/a)$, among $n$ values of $p_{ka, i}^{(\sigma)}(=0,1)$ ($i=1, \cdots, n$),
the number of $p_{ka, i}^{(\sigma)}=1$ is $|m_1-m_2|$, which is equal to the number of times the sequence $\left( \cos{(\phi)}, \cdots, \cos{\left(\phi+n\xi\right)} \right)$ (or $\left( \sin{(\phi)}, \cdots, \sin{\left(\phi+n\xi\right)} \right)$) changes sign.
One can easily check this conclusion in Table \ref{tab:phase}.
For example, the anticrossing at $k=0$, 
$\omega \sim 2$ in Fig.~\ref{fig:phase} (b-2) is between the bands with $\sigma = 2, 3$, having $m=\pm 1$.
It corresponds to $m_1=1$ and $m_2=-1$ leading to $|m_1 - m_2|=2$.
Thus, among the three values of $p_{0, i}^{(2)}$, two (i.e. $i=1$ and $i=2$) have the value of $1$, in accordance with Table \ref{tab:phase}. 
The similar conclusion holds also for $p_{0, i}^{(3)}$. {
We explain another example on the anticrossing at $k=\pi$, 
$\omega \sim 1.1$ in Fig.~\ref{fig:phase} (b-2) between the bands with $\sigma = 1$, and $\sigma=2$, corresponding to $m_1=0$ and $m_2=\pm 1$ i.e. with $|m_1 - m_2|=1$.
Thus, among the three values of $p_{\pi, i}^{(1)}$ (and similarly among the three values of $p_{\pi,i}^{(2)}$), only one has the value of $1$, 
in accordance with Table \ref{tab:phase}. }

To confirm the theory in this section, we calculate the relative phase $\theta(k)$ for the material $\ce{Li6B5}$ with approximate translational symmetry $\hat{T}_{a/5}$, which are shown in Sec.~\ref{sec:first-principles calculation}.

\section{\label{sec:3D helix} Phonons in 3D systems with approximate screw symmetry}

In this section, {we extend the result in Sec.~\ref{sec:1D chain} for 1D systems with approximate translation symmetry to 3D systems with approximate screw symmetry.}
In such 3D systems, the pseudoangular momentum $m$ plays the role of the quantum number in Sec.~\ref{sec:1D chain}.
Namely, we discuss a method to extract the information of the pseudoangular momentum by calculating the relative phases in phonon systems with approximate screw symmetry in 3D systems.

We suppose the system has the $n_l$ screw symmetry $C_n^l$ in the $z$ direction, where $n$ and $l$ are positive integers with $1 \le l < n$.
The screw operation $C_n^l$ is a combination of a rotation by an angle $2\pi/n$ around the $z$-axis, 
and a translation by $(l/n)a$ along the $z$ axis, where $a$ denotes the lattice constant along the $z$ axis. 
Then its $n$th power is equal to the the translation $T_{la}$ by $la$: $\hat{T}_{l a} = \left( \hat{C}_n^l \right)^n$.
From the Bloch's theorem, an eigenfunction $\psi(x)$ satisfies $\hat{T}_{a} \psi(x) = e^{-ika} \psi(x)$, where $k$ is the Bloch wavenumber $\left( \frac{-\pi}{a} < k \le \frac{\pi}{a} \right)$ in the $z$ direction. Since we need to consider eigenstates under $\hat{C}_n^l$, 
the wavevector $\vec{k}=(k_x,k_y,k_z)$ should be invariant under this screw symmetry. Here we choose $(k_x, k_y)=(0,0)$, and write 
$\vec{k}=(0,0,k)$. 
Therefore, when the $n_l$ screw symmetry $\hat{C}_n^l$ is preserved, the eigenfunction $\psi(x)$ can be characterized by a integer $m'$ and satisfies $\hat{C}_n^l \psi(\vec{r}) = e^{i(kla+2m'\pi)/n} \psi(\vec{r})$, where 
\begin{align}
  m' \equiv \left\{\begin{array}{ll}
    0, \pm 1, \cdots, \pm (n-1)/2 & \text{for} \ n \ \text{odd} \\
    0, \pm 1, \cdots, \pm (n-2)/2, n/2 & \text{for} \ n \ \text{even}
  \end{array}\right. \pmod n .
\end{align}
In the following, we call $m'$ a pseudoangular momentum.
We note that this $m'$ corresponds to the quantized integer part of the pseudoangular momentum defined in Ref.~\cite{zhang2021chiral}.

Next, we extend our theory in the previous section to 3D systems with approximate screw symmetry, which needs to define relative phases between neighboring atoms.
In this section, we restrict ourselves to the case where $n$ and $l$ are coprime for simplicity.
The cases where $n$ and $l$ are not coprime can be studied similarly, as shown in Appendix~\ref{app:not coprime}.
When $n$ and $l$ are coprime, the $n$ atoms in the unit cell are {periodically located with a period $a/n$ along the screw axis}, which we call $z$ axis.
Now, we show that there exists a symmetry operation 
$\hat{\mathcal{O}}\equiv (\hat{C}_n^l)^{p} (\hat{T}_{a})^{q}$ ($p$, $q$: integer) which relates between neighboring atoms.
Since the difference in the $z$-component between the neighboring atoms is $\frac{a}{n}$, the integers $p$ and $q$ satisfy $p \frac{la}{n} + q a = \frac{a}{n}$, i.e. $l p + n q = 1$. Thanks to the B\'{e}zout's lemma this equation always has integer solutions for $p$ and $q$.
Specifically, for $l=2$ and $n=7$, this equation has a solution $p=-3$ and $q=1$.
Hence $\hat{\mathcal{O}}=(\hat{C}_n^l)^{p} (\hat{T}_{a})^{q}$ is the operator that rotates the system by $\frac{2 p \pi}{n}$ and translates it by $a/n$.
Therefore, 
\begin{align}
\hat{\mathcal{O}}\psi(\vec{r})&= (\hat{C}_n^l)^{p} (\hat{T}_{a})^{q} \psi(\vec{r}) \notag \\
  &= \left(e^{-i (lka + 2m'\pi)/n}\right)^{p} \left(e^{i k a}\right)^{q} \psi(\vec{r}) 
  \label{eq:Opsi}
\end{align}
holds. Then, similarly to the definition of the pseudoangular momentum, 
we put the eigenvalue of $\hat{\mathcal{O}}$ to be equal to $ e^{i (ka + 2m\pi)/n}$ because $\hat{\mathcal{O}}^n=e^{ika}$, where $m$ 
is an integer representing the relative phase between neighboring atoms displaced by $a/n$: 
\begin{align}
	\hat{\mathcal{O}}\psi(\vec{r})=e^{-i (ka + 2m\pi)/n}\psi(\vec{r}).
	  \label{eq:Opsi2}
\end{align}
Then, from (\ref{eq:Opsi}) and (\ref{eq:Opsi2}), $m$ is 
given by $m \equiv m' p \pmod{n}$.
By using $lp+nq=1$, we get 
\begin{align}
&m' \equiv ml \pmod{n}.
\label{eq:mml}
\end{align}
In particular for $n_1$ screw symmetry (i.e.~$l=1$), $m' \equiv m$ holds.
Therefore, from this correspondence, the wave function information possessed by the integers $m$ and $m'$ is the same.
In the following discussion,
we consider assigning the integer $m$ instead of the pseudoangular momentum $m'$ {for phonon} eigenmodes.

Let $\bm{\epsilon}^{(\sigma)}(k)= (\vec{\epsilon}_1^{(\sigma)}(k), \cdots, \vec{\epsilon}_n^{(\sigma)}(k))^{\mathrm{T}}$ denote the eigenvector of the dynamical matrix $D(\vec{k})$, where $\vec{\epsilon}_j^{(\sigma)}(k)$ is the displacement of the $j$-th atom in the $\sigma$-th phonon mode at the wavenumber $k$ along the $z$ axis. Here the $n$ atoms are numbered in the increasing order of the $z$ coordinate. 
This means that the eigenvector of the dynamical matrix $D(k)$ with the screw symmetry $C_n^l$ can be expressed in terms of the pseudo angular momentum $m$ as
\begin{align}
  \label{eq:exact screw eigenvector}
  \bm{\epsilon}^{(\sigma)}(k) &= \left(\begin{array}{c}
    \vec{v}^{(\sigma)}(k)  \\
    e^{i(ka+2m\pi)/n} A_{p} \vec{v}^{(\sigma)}(k) \\
    \vdots \\
    e^{i(n-1)(ka+2m\pi)/n} A_{p}^{n-1} \vec{v}^{(\sigma)}(k)
  \end{array}\right) , \\
  A_{p} &= \left(\begin{array}{ccc}
    \cos{2p\pi/n} & -\sin{2p\pi/n} & 0 \\
    \sin{2p\pi/n} & \cos{2p\pi/n} & 0 \\
    0 & 0 & 1
  \end{array}\right) ,\label{eq:Ap}
\end{align}
where $\vec{v}^{(\sigma)}(k)$ is a three-dimensional complex vector.
Thus, when the system has exact screw symmetry, the eigenvector is characterized by the quantum number $m$.
On the other hand, without the screw symmetry, we cannot define the quantum number $m$ for the eigenvector.
However, in the case with approximate screw symmetry, the information of the quantum number $m$ defined under exact screw symmetry partially remains in the eigenvector.

Based on Eq.~(\ref{eq:Opsi2}), under the exact screw symmetry, we define the relative phase to be
\begin{align}
  z_{i, \alpha}^{(\sigma)} (k) &= \left\{ \begin{array}{ll}
    u_{i+1, \alpha}^{(\sigma)}(k) / u_{i, \alpha}^{(\sigma)}(k) & i = 1, \cdots, n-1 \\
    u_{1, \alpha}^{(\sigma)}(k) e^{ika} / u_{n, \alpha}^{(\sigma)}(k) & i = n
  \end{array} \right. , \\
  \theta_{i, \alpha}^{(\sigma)}(k) &= \arg z_{i, \alpha}^{(\sigma)} (k) ,
\end{align}
for the $\sigma$-th eigenvectors $\bm{\epsilon}^{(\sigma)}(k)$ of the dynamical matrix $D(k)$,
where $\alpha = x, y, z$ and $\vec{u}_{i}^{(\sigma)}(k) = A_{p}^{-(i-1)}\vec{\epsilon}_{i}^{(\sigma)}(k),\ (i=1, 2, \cdots, n)$.
In the case of the exact screw symmetry, we can use Eq.~(\ref{eq:exact screw eigenvector}) to calculate the relative phase as 
\begin{align}
&\theta_{i, \alpha}^{(\sigma)} (k) = (ka + 2m\pi)/n, (i=1,\cdots,n \ ,\alpha = x, y, z);
\label{eq:thetascrew}
\end{align}
namely, 
$\theta_{i, \alpha}^{(\sigma)} (k)$ is a linear function with an intercept of $2m\pi/n$ independent of the direction $\alpha$.

\begin{figure}
  \includegraphics[clip,width=\linewidth]{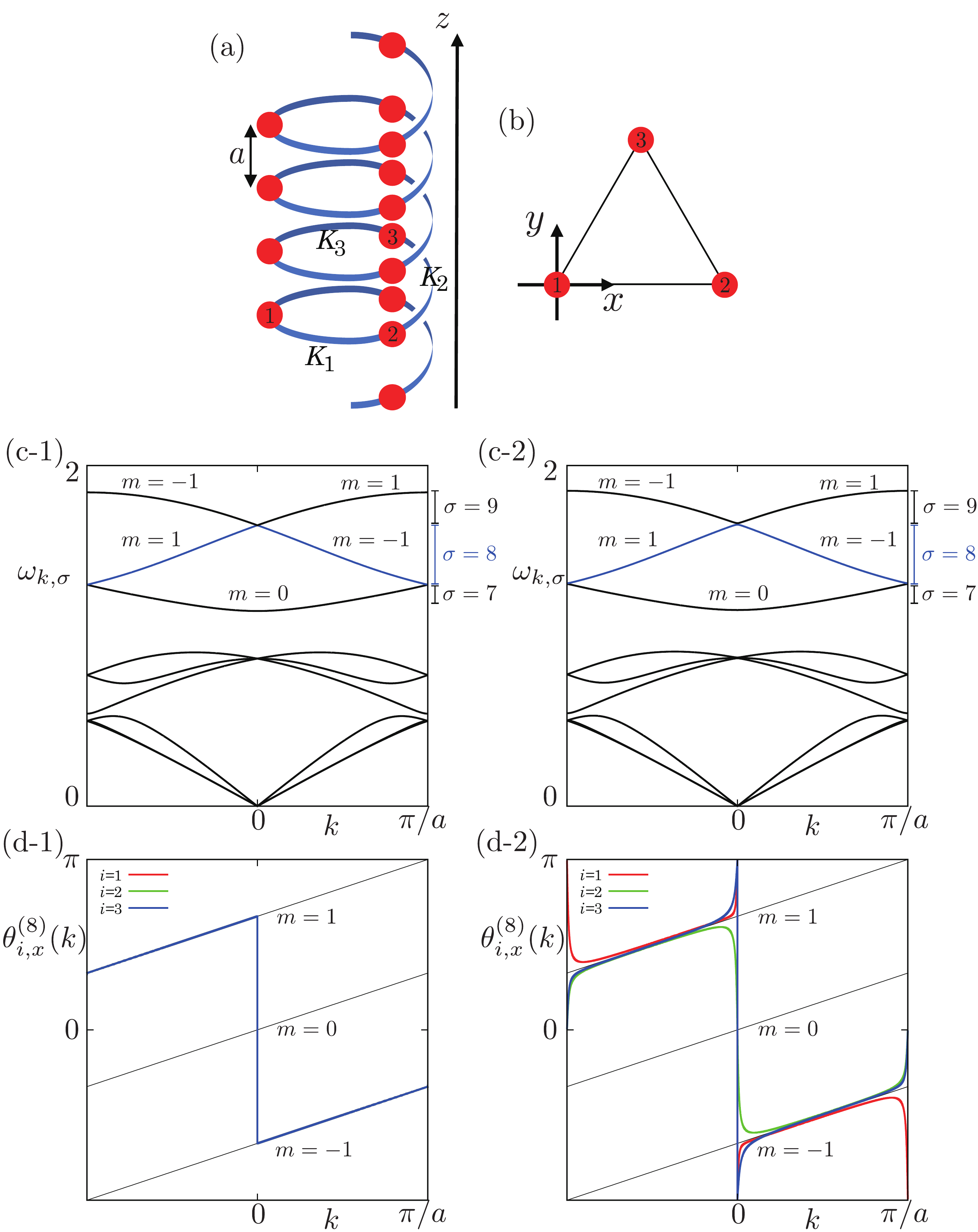}
  \caption{\label{fig:screw_model_phase} 
  Phonon system with three-fold $3_1$ screw symmetry in the $z$ direction.
  (a), (b) Schematic picture of this model, and red dots denote atoms.
  (a) shows that this is the $3_1$ screw symmetry.
  (b) shows the positions of the atoms within the $xy$ plane.
  (c-1) (c-2) Band dispersions $\omega_{k, \sigma}$ with (c-1) exact screw symmetry and (c-2) broken screw symmetry.
  (d-1) (d-2) Relative phases $\theta_{i, x}^{\sigma} (k)$ for the $\sigma=8$ band (blue line) in the band dispersions (c-1) and (c-2).
  (d-1) corresponds to the case with exact screw symmetry, and (d-2) corresponds to the case with broken screw symmetry.
  }
\end{figure}
As an example, we calculate phonon modes in a model that slightly breaks  threefold screw symmetry $C^1_3$. Here, because $n=3$ and $l=1$, the quantum number $m$ is equal to the pseudoangular momentum $m'$.
We consider a system with three atoms 1, 2 and 3 in a unit cell shown in Figs.~\ref{fig:screw_model_phase} (a) {and} (b), arranged along a helix.
The result of the phonon calculation is shown in Figs.~\ref{fig:screw_model_phase} (c) {and} (d).
In this calculation, let $K_1$, $K_2$ and $K_3$ denote the spring constant matrices between atom $1$ and atom $2$, between atom $2$ and atom $3$, and between atom $3$ and atom $1$ of the neighboring unit cell, respectively.
The spring constant matrix $K_{i}$ consists of longitudinal components with a spring constant $K_{L,i}=c_{i}$ and transverse ones with a spring constant $K_{T,i}=c_{i}/4$.
In our calculation, we change the parameter values from those with the exact screw symmetry $\left( c_1=c_2=c_3=1 \right)$ to those with approximate screw symmetry $\left( c_1=1, c_2=1.01, c_3=1.02 \right)$.
When the screw symmetry is slightly broken in this way, the band dispersion relation changes slightly from Fig.~\ref{fig:screw_model_phase} (c-1) to (c-2), with a tiny gap opened at $k=0$.
Here, if we take the $\sigma = 8$ band (blue lines in (c-1) and (c-2)) as an example, the relative phase change from Fig.~\ref{fig:screw_model_phase} (d-1) to (d-2).
In the case of exact screw symmetry, the relative phase is a linear function, Eq.~(\ref{eq:thetascrew}).
Since the value of the relative phase $\theta_{i, \alpha}^{(\sigma)}(k)$ {hardly changes even when} the screw symmetry is slightly broken, we can use this relative phase to assign the pseudoangular momentum $m(=m')$ to all the bands, except for the vicinity of band anticrossing.
As shown in Fig.~\ref{fig:screw_model_phase} (c-1), 
when the wavenumber $k$ along the $z$ axis increases across the Brillouin zone boundary from $k=\pi/a$ to $k=-\pi/a$, the quantum number $m \pmod{n}$ increases by $+1$.
This result is similar to that of the one-dimensional model in Fig.~\ref{fig:phase}, and can be interpreted by using the extended Brillouin zone {method} \cite{zhang2021chiral}.
In addition, the relative phase $\theta_{i, \alpha}^{(\sigma)}(k)$ with broken symmetry is significantly different from that with exact symmetry near the time-reversal invariant momenta $k=0, \pi/a$, ant it is quantized to $0$ or $\pi$ on $k=0, \pi/a$ because its eigenvector is real.
In the one-dimensional system with broken $T_{a/n}$ translation symmetry (Fig.~\ref{fig:phase}), the property that the number of $\theta_{i, \alpha}^{(\sigma)}(k)$ ($i=1,\cdots,n$) at $k=0, \pi/a$ going to $\pi$ is equal to the difference $|m_1 - m_2|$ of pseudoangular momentum $m = m_1, m_2$ of the crossing bands holds as in the one-dimensional case.
However, this property is not generally valid in three-dimensional systems with slightly broken screw symmetry.

\section{\label{sec:first-principles calculation} First-principles calculation}
In this section, we calculate the relative phases between atoms in phonon modes in a material $\ce{Li6B5}$\cite{wang2017prediction} with approximate translational symmetry (discussed in Sec.~\ref{sec:1D chain}) 
and two materials $\ce{SnIP}$ \cite{pfister2016inorganic} and $\ce{S10}$ \cite{lind1969structure} with approximate screw symmetries by first-principle calculations (discussed in Sec.~\ref{sec:3D helix}).
By comparing their relative phases with the ones with exact symmetry, we can extract the information of the quantum number $m$.
$\ce{Li6B5}$, $\ce{SnIP}$ and $\ce{S10}$ are fully relaxed before the phonon calculations.
We calculate the force constants of 
$\ce{Li6B5}$, $\ce{SnIP}$ and $\ce{S10}$ with density functional perturbation theory (DFPT) in a $3\times3\times3$, $2\times2\times2$ and $2\times2\times3$ supercell with an equivalent $k$-mesh of $9\times9\times9$, $6\times6\times4$ and $4\times4\times6$ via VASP \cite{CAL_VASP, CAL_DFPT}, respectively.

\subsection{Li6B5: one-dimensional chain with approximate translational symmetry}
\begin{figure}
  \includegraphics[clip,width=\linewidth]{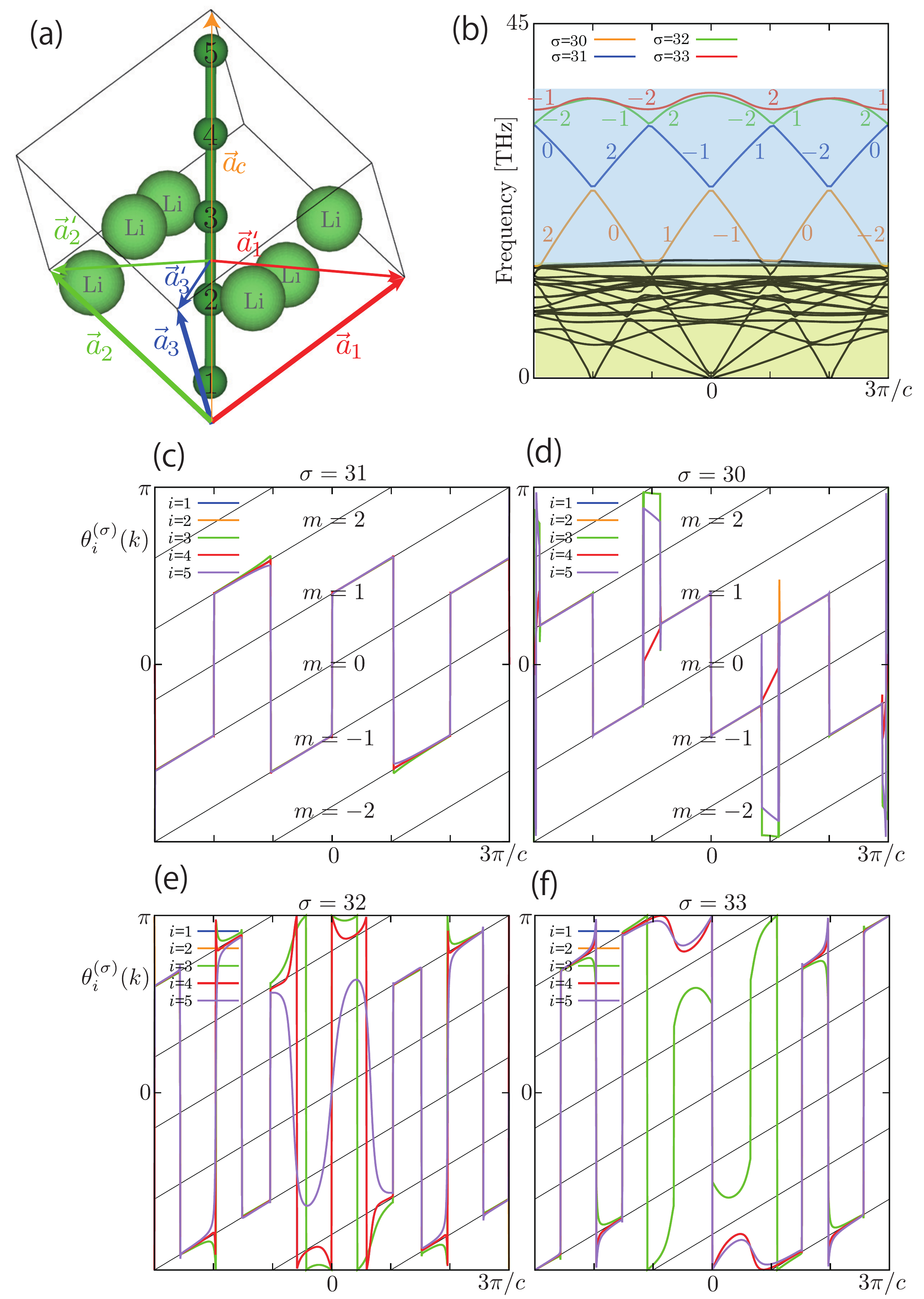}
  \caption{\label{fig:Li6B5} 
  Relative phase of the  material $\ce{Li6B5}$.
  (a) Schematic picture of the atomic positions.
  The boron atoms have the approximate translational symmetry $T_{c/5}$.
  (b) Phonon band dispersion.
  The bands ($\sigma=30, 31, 32, 33$) shown in color in the figure are phonon modes with displacements along the direction $\vec{a}_c = \vec{a}_1 + \vec{a}_2 + \vec{a}_3$.
 {The band structure is classified into two groups, the one with a wider bandwidth ($\sim 40$THz, shown in the light blue background) and the other with a narrower bandwidth ($\sim 15$THz, shown in the light yellow background).
  (c)-(f) Relative phases between neighboring atoms of the bands $\sigma=$ (c) 31, (d) 30, (e) 32 and (f) 33, respectively.
  }}
\end{figure}
We discuss the results of the calculations for the material $\ce{Li6B5}$ shown in Fig.~\ref{fig:Li6B5}.
{$\ce{Li6B5}$ contains a one-dimensional chain of boron atoms with approximate translational symmetry $\hat{T}_{c/5}$, where $c$ is the lattice constant along the $z$ axis, as shown in Fig.~\ref{fig:Li6B5} (a).}
The lattice is rhombohedral and its lattice vectors are 
\begin{align}
	&\vec{a}_1 = (\sqrt{3}a/2, a/2, c/3),\\&\vec{a}_2 = (-\sqrt{3}a/2, a/2, c/3),\\ &\vec{a}_3 = (0, -a, c/3),
	\end{align}
where $a= 3.973$\AA\ and $c = 7.945$\AA. The corresponding primitive reciprocal vectors are
\begin{align}
	&\vec{b}_1 = (2\pi/(\sqrt{3}a), 2\pi/(3a), 2\pi/c) ,\\
	&\vec{b}_2 = (-2\pi/(\sqrt{3}a), 2\pi/(3a), 2\pi/c),\\
	&\vec{b}_3 = (0, -4\pi/3a, 2\pi/c).
\end{align}
The unit cell contains five boron atoms $\ce{B}$ almost equally spaced along the vector $\vec{a}_{c} = \vec{a}_1 + \vec{a}_2 + \vec{a}_3 = (0, 0, c)$.
Thus, in this material the boron atoms have an approximate translational symmetry $\hat{T}_{c/5}$ by the vector $(0, 0, c/5)$.
Since this lattice has a translational symmetry $\hat{T}_{i}$ $(i = 1,2,3)$ in the direction of the lattice vectors $\vec{a}_i$, the eigenfunctions $\psi(x)$ satisfy $\hat{T}_{i} \psi(x) = e^{-i \vec{k}\cdot \vec{a}_i} \psi(x)$ where $-\pi < \vec{k} \cdot \vec{a}_i \le \pi$ due to the Bloch's theorem.
By using this, under the translation symmetry $\hat{T}_a$ in the direction of $\vec{a}_{c}$, $\hat{T}_{c}= \hat{T}_1 \hat{T}_2 \hat{T}_3$, $\psi(x)$ satisfies
\begin{align}
  \hat{T}_{c} \psi(x) = e^{-i \vec{k} \cdot \vec{a}_c} \psi(x) ,\ \left( -3 \pi < \vec{k} \cdot \vec{a}_c \le 3\pi \right) .
\end{align}
Furthermore, if the boron atoms preserve the translational symmetry $\hat{T}_{c/5}$, 
\begin{align}
  \hat{T}_{c/5} \psi(x) = e^{-i (\vec{k} \cdot \vec{a}_c + 2 m \pi )/5} \psi(x)
\end{align}
holds with  the quantum number $m = 0, \pm 1, \pm 2$.

Figure~\ref{fig:Li6B5} (b) shows the phonon band structure along the $k_z$ direction.
The band structure is classified into two groups, the one with a wider bandwidth ($\sim 40$THz) and the other with a narrower bandwidth ($\sim 15$THz).
These  two groups of bands correspond to the phonon modes with displacements along $z$ axis and these along the $xy$ plane, respectively.
In the former group with displacements along $z$ axis, the motions of the $\ce{B}$ atoms are well decoupled from those of the $\ce{Li}$ atoms, which is ideal for our purpose. Henceforth, we consider the bands $\sigma=30, 31, 32, 33$, having boron atoms vibrating only in the $z$ direction.
Thus, when $\hat{T}_{a/5}$ translation symmetry is preserved, we can write the eigenvector $\bm{\epsilon}_{k}^{(\sigma)}$ for the phonon eigenvectors with displacements along the $z$ direction under this assumption, as 
\begin{align}
  \bm{\epsilon}_{k}^{(\sigma)} = \left(\begin{array}{c}
    1 \\
    e^{i (\vec{k} \cdot \vec{a}_c + 2 m \pi )/5} \\
    e^{2i (\vec{k} \cdot \vec{a}_c + 2 m \pi )/5} \\
    e^{3i (\vec{k} \cdot \vec{a}_c + 2 m \pi )/5} \\
    e^{4i (\vec{k} \cdot \vec{a}_c + 2 m \pi )/5}
  \end{array}\right) , 
\end{align}
where $m$ is the quantum number $m = 0, \pm 1, \pm 2 \pmod{5}$.
Here $\vec{k}\cdot \vec{a}_c$ takes values within the range $-3\pi < \vec{k}\cdot \vec{a}_c \le 3\pi$, so the wavenumber $k = \vec{k}\cdot \vec{a}_c / c$ in the $z$ direction takes the range $\frac{-3\pi}{c} < k \le \frac{-3\pi}{c}$.
Therefore, as we showed in Sec.~\ref{sec:1D chain}, the relative phase from this eigenvector is
\begin{align}
  \label{eq:Li6B5 phase}
  \theta(\vec{k}) = (k c + 2 m \pi )/5 .
\end{align}
{However, the translational symmetry ${T}_{c/5}$ is slightly broken in this material}, so the relative phase is slightly deviated from Eq.~(\ref{eq:Li6B5 phase}). 
{But} the deviation is small, and we can still extract the information of the quantum number $m$.
In particular, the relative phase of the band $\sigma = 31$ is shown in Fig.~\ref{fig:Li6B5} (c), which agrees well with the relative phase of the case with exact symmetry (Eq.~(\ref{eq:Li6B5 phase})). Thus, we can extract the quantum number $m$ as shown in the figure.
The results of the relative phase calculations for the bands $\sigma = 30,32,34$ are shown in Fig.~\ref{fig:Li6B5} (d)-(f).
Since the relative phase of $\sigma=30$ agrees well with the relative phase of the case with exact symmetry, we can extract the quantum number $m$.
On the other hand, the relative phases of $\sigma=32$ and $33$ do not match those of the exact symmetric case near $k=0$.
This is because the bands are close to each other and thus hybridized. The results for the quantum number $m$ is summarized in in Fig.~\ref{fig:Li6B5}(b). The behavior of $m$ for the bands $\sigma=32,33$ will be discussed later.
.

\begin{figure*}
  \includegraphics[clip,width=\linewidth]{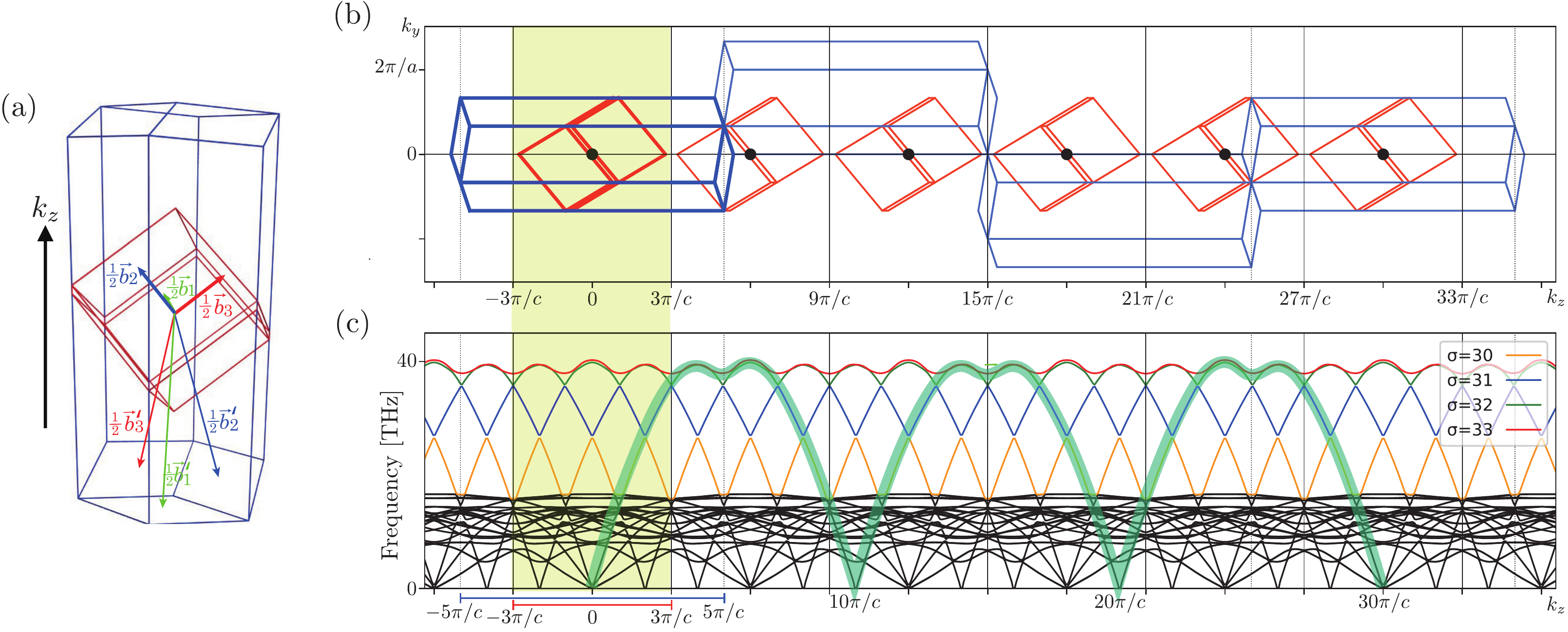}
  \caption{\label{fig:extend Li6B5} 
  Extended Brillouin zone of $\ce{Li6B5}$.
  (a) Comparison of Brillouin zones with and without the translational symmetry $\hat{T}_{c/5}$.
  The blue line is the virtual Brillouin zone with $\hat{T}_{c/5}$, and the red line is the actual Brillouin zone without $\hat{T}_{c/5}$.
  (b) Extended Brillouin zone with actual Brillouin zones on $k_z$ axis in the reciprocal lattice space.
  (c) Band dispersion on $k_z$ axis. The band shown by the thick green line is the phonon mode with the quantum number $m=0$.
  }
\end{figure*}
We can assign the quantum number $m$ from the relative phases as shown in Fig.~\ref{fig:Li6B5} (b).
In this figure, when the wavenumber increases across the boundary of the Brillouin zone from $k=3\pi/c$ to $k=-3\pi/c$, the quantum number $m \pmod{5}$ increases by $+3$.
This is quite different from the increase by $+1$ across the boundary of the Brillouin zone in Fig.~\ref{fig:phase}.
In the following, we explain this difference using the extended band scheme shown in Fig.~\ref{fig:extend Li6B5}.

Let us assume that the system is invariant under the translation by the vector $\vec{a}_c/5=(0,0,c/5)$, meaning that the 
five atoms within the unit cell are equivalent under this symmetry. Thus, the
new primitive translation vectors  become 
\begin{align}
&\vec{a}'_1 = \vec{a}_1 - 2 \vec{a}_c/5 = (\sqrt{3}a/2, a/2, -c/15), \\
&\vec{a}'_2 = \vec{a}_2 - 2 \vec{a}_c/5 = (\sqrt{3}a/2, a/2, -c/15),\\
&\vec{a}'_3 = \vec{a}_3 - 2 \vec{a}_c/5 = (0, -a, -c/15). \end{align}
Their sum is equal to $-\vec{a}_c/5$, representing the $\hat{T}_{c/5}$ translation symmetry. Because the translation vector along the $z$ axis becomes shorter, the 
Brillouin zone becomes longer. These tranlation vectors lead to 
a virtual Brillouin zone given by corresponding primitive reciprocal vectors
\begin{align}
&\vec{b}'_1 = (2\pi/(\sqrt{3}a), 2\pi/(3a), -10\pi/c) ,\\
&\vec{b}'_2 = (-2\pi/(\sqrt{3}a), 2\pi/(3a), -10\pi/c),\\
&\vec{b}'_3 = (0, -4\pi/3a, -10\pi/c).
\end{align}
Figure~\ref{fig:extend Li6B5} (a) is a figure comparing the virtual Brillouin zone (blue line) when the system is assumed to have translational symmetry $\hat{T}_{c/5}$ and the actual Brillouin zone (red lines) when it is not assumed.
These Brillouin zones are of the similar shape, consisting of 12 faces. 
Namely, the virtual Brillouin zone (blue lines) for the case with the translational symmetry $\hat{T}_{c/5}$ is narrowed to the actual Brillouin zone (red line) by breaking the symmetry.
Figure~\ref{fig:extend Li6B5} (b) shows five virtual Brillouin zones along $k_z$ axis and the accompanying actual Brillouin zones in Fig.~\ref{fig:extend Li6B5} (a).
Furthermore, we show the band dispersion relation along the $k_z$ axis in Fig.~\ref{fig:extend Li6B5} (c), to compare with the Brillouin zones in Fig.~\ref{fig:extend Li6B5} (b).
As a one-dimensional system, the translation vector $(0,0,c/5)$, gives rise to the one-dimensional reciprocal vector $10\pi/c$, 
and the complex structure of the colored bands in Fig.~\ref{fig:Li6B5}(b) is then mapped to a single band in the extended BZ with the size $10\pi/c$, shown as
the light green line in Fig.~\ref{fig:extend Li6B5} (c). This single band corresponds to the  quantum number $m=0$, as we discussed in Sec.~\ref{sec:1D chain}.
{In this case, the size of the actual Brillouin zone along $k_z$ is $6\pi/c$ because $\vec{b}_{c} = \vec{b}_1 + \vec{b}_2 + \vec{b}_3 = (0, 0, 6\pi/c)$,
and 
the quantum number $m \pmod{5}$ increases by $+3$ from Eq.~(\ref{eq:Li6B5 phase}) if we shift the wavenumber by $-6\pi/c$.} In this way, in folding the virtural BZ to the actual BZ, 
if the phonon bands are shifted by  $(-6\pi/c)\times \bar{m}$, ($\bar{m}$: integer), the quantum number $m$ characterizing the relative phases between neighboring atoms, 
is equal to $\bar{m}$. The value of $m$ for the phonon bands completely agrees with that obtained in Fig.~\ref{fig:Li6B5} (b). Thus the phonon wavefunctions 
retains information for the case with $T_{c/5}$ symmetry. 

{Now we discuss in which cases the quantum number $m$ is well defined and the screw symmetry is regarded as a relatively good symmetry. 
In Figs.~\ref{fig:Li6B5}(c)-(f), we can compare the result for the quantum number $m$ for the bands $\sigma=30,31,32,33$. We see that the quantum number $m$ is well defined in most cases, except for the bands $\sigma=32,33$ near $k=0$ (Fig.~\ref{fig:Li6B5}(e)-(f)). From the phonon dispersion (Fig.~\ref{fig:Li6B5}(b)) the bands $\sigma=32, 33$ are close to each other, making them to hybridize and to make the quantum number $m$ relatively ill-defined. One may wonder that the bands $\sigma=32, 33$ are mutually close not only near $k=0$ but also around $k=\pm 2\pi/c$, but even near  $k=\pm 2\pi/c$ the quantum number $m$ is well defined for 
$\sigma=32, 33$ as seen in Fig.~\ref{fig:Li6B5}(e)-(f). One can understand 
this difference between  $k\sim 0$ and $k\sim \pm 2\pi/c$ by the above scheme of the virtual Brillouin zone.
As mentioned in the present subsection, the size of the reciprocal lattice vector along the $c$-axis is $6\pi/c$, which gives rise to the folding of the Brillouin zone. This folding originates from the Umklapp scattering of
phonons by the wavenumber $\pm 6\pi/c$. 
From Fig.~\ref{fig:extend Li6B5} (b), the near degeneracy at $k\sim 0$ between the $\sigma=32, 33$ bands is between the bands originally at $k\sim 6\pi/c$ and $k\sim -6\pi/c$, via the second-order effect in the 
Umklapp scattering. 
On the other hand, 
 the near degeneracy at $k\sim 2\pi/c$ between the bands $\sigma=32, 33$ is between the bands originally at $k\sim 14\pi/c$ and $k\sim -4\pi/c$, via the third-order effect in the 
Umklapp scattering. The near degeneracy at $k\sim -2\pi/c$ is treated similarly as the third order in the Umklapp scattering. Thus the near degeneracy around $k\sim \pm 2\pi/c$ is of the higher order in the Umklapp scattering than that around $k\sim 0$, and the hybridization is weaker. It is the reason why the bands $\sigma=32,33$ 
near $k\sim \pm 2\pi/c$ do not hybridize so much (and the quantum number $m$ is well defined), compared with those near $k\sim 0$, in agreement with the result in Fig.~\ref{fig:Li6B5}(e)-(f). }

\subsection{SnIP and S10: materials with approximate screw symmetry}
\begin{figure}
  \includegraphics[clip,width=\linewidth]{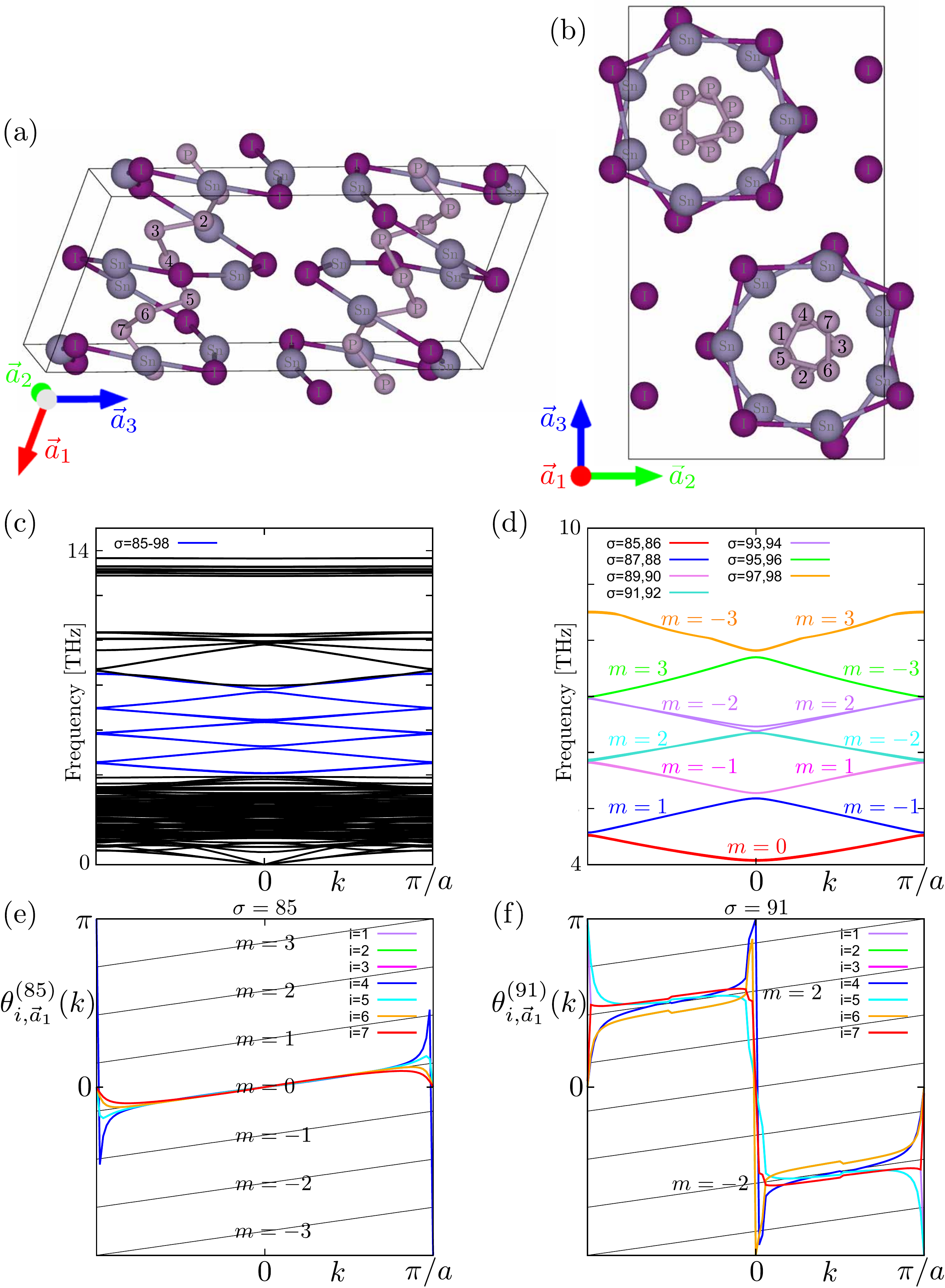}
  \caption{\label{fig:SnIP}
  Approximate screw symmetry of phonons in the material $\ce{SnIP}$
  (a), (b) Schematic pictures of atomic positions. 
  (a) show that one helical chains has approximate $7_2$ screw symmetry in the $\vec{a}_1$ direction.
  (b) shows the atomic positions as seen from the $\vec{a}_1$ direction.
  (c) (d) Band dispersions.
  In (c) we highlight the focused bands ($\sigma=85,\cdots,92$) in blue.
  The bands with $\sigma=85, \cdots ,92$. are shown in (d).
  We show the value of the quantum number $m$ for each band from the relative phase discussion.
  (e) (f) Relative phases of the bands $\sigma = 85$ and $91$, respectively.
  }
\end{figure}
First, we explain the calculations for $\ce{SnIP}$ shown in Fig.~\ref{fig:SnIP}.
The atomic position in the unit cell is shown in Figs.~\ref{fig:SnIP} (a)-(b), where the unit cell has two helical chains with opposite chiralities and $42$ atoms.
One helical chain has approximate $7_2$ screw symmetry shown in Fig.~\ref{fig:SnIP} (b), and this chain is transformed to the other one by glide reflection.
Figure~\ref{fig:SnIP} (c) shows the phonon dispersions in $\ce{SnIP}$, which has $126$ bands.
In this paper, we analyze the approximate screw symmetry of the helical chains formed by $\ce{P}$ atoms, and we focus on the bands with indices $\sigma = 85, \cdots, 98$ shown in Fig.~\ref{fig:SnIP} (d).
All of these bands ($\sigma=85, \cdots, 98$) are almost doubly degenerate because phonon modes in the two helical chains in the unit cell are almost decoupled, and 
these two chains have the same phonon eigenfrequencies due to the glide symmetry \cite{Peng-SnIP-2021}.
Therefore, we estimate the quantum number $m$ by extracting one of the helical chains and calculating the relative phase.
Figures~\ref{fig:SnIP} (e) and (f) show the calculations of the relative phases for the $\sigma=85$ and $\sigma = 91$ bands, respectively.
The calculations for the other bands are given in Appendix~\ref{app:other band}.
By comparing them with the relative phases for the case with exact screw symmetry (Eq.~(\ref{eq:Li6B5 phase})), we can assign the value of the quantum number $m$ for each band as shown in Fig.~\ref{fig:SnIP} (d).
The value of the pseudoangular momentum $m'$ (associated with $7_2$ symmetry) is calculated by $m' \equiv 2 m \pmod{7}$, 
{which follows from Eq.~(\ref{eq:mml}) for the case with mutually coprime $n$ and $l$.}
We can see that the quantum number $m$ of the band $\sigma=85$ is $0$ and that of the band $\sigma = 91$ changes from $2$ to $-2$ at the $\Gamma$ point.
Similarly, we can assign the values of $m$ for each band as presented in Appendix \ref{app:other band} without ambiguity, and 
the result is shown in Fig.~\ref{fig:SnIP} (d).
In agreement with our discussion in Sec.~\ref{sec:3D helix}, the quantum number $m \pmod{7}$ increases by $+1$ across the Brillouin zone boundary.
Thus, even when the $7_2$ screw symmetry is broken {in SnIP}, the eigenvectors of the system still have properties that well reflect the approximate screw symmetry.

\begin{figure}
  \includegraphics[clip,width=\linewidth]{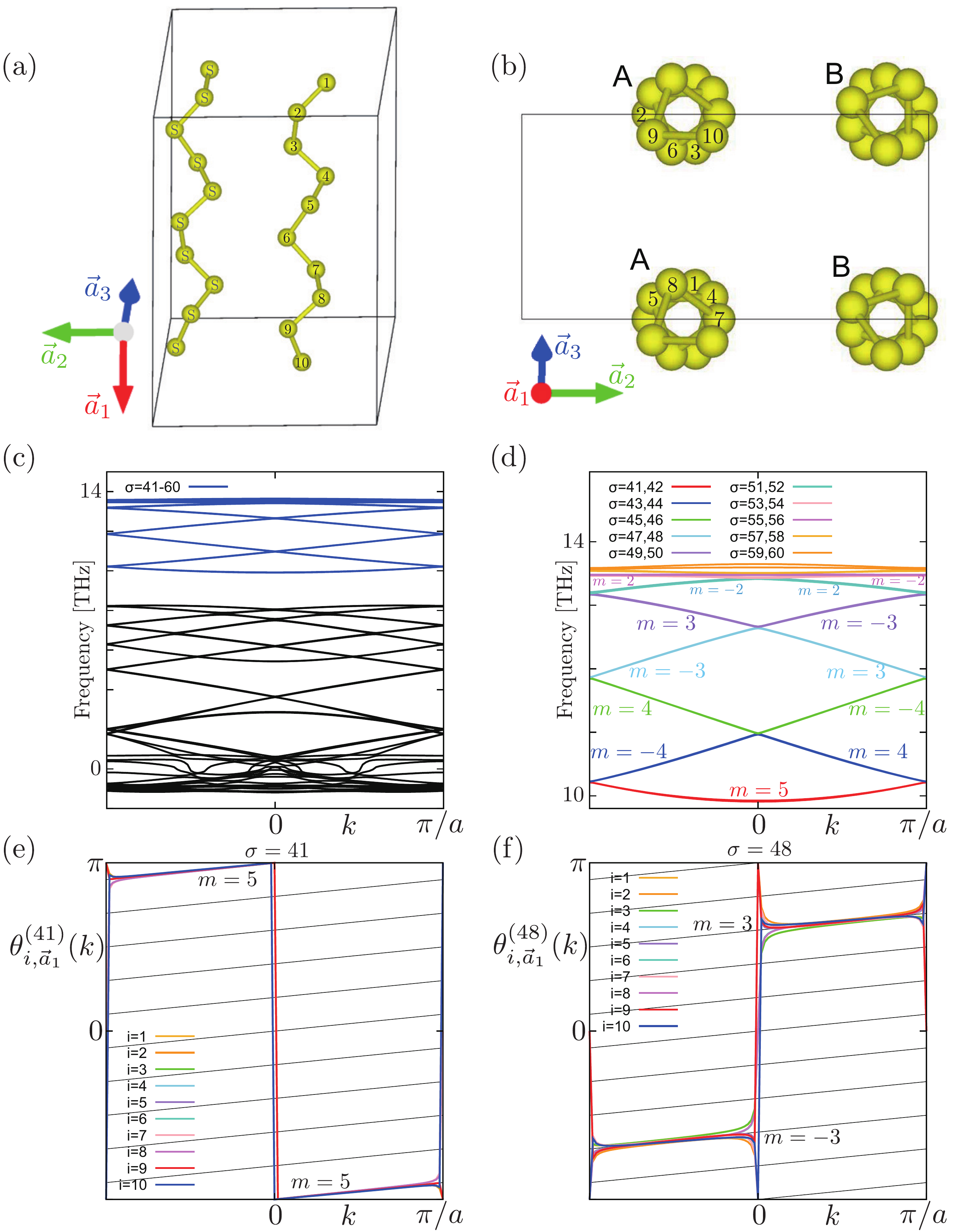}
  \caption{\label{fig:C10}
  Approximate screw symmetry of phonons in the material $\ce{S10}$
  (a), (b) Schematic pictures of atomic positions. 
  (a) show that one helical chains has approximate $10_7$ screw symmetry in the $\vec{a}_1$ direction.
  (b) shows the atomic positions as seen from the $\vec{a}_1$ direction.
  (c) (d) Band dispersions.
  In (c) we highlight the focused bands ($\sigma=41,\cdots,60$) in blue.
  The bands with $\sigma=41, \cdots ,60$. are shown in (d).
  We show the value of the quantum number $m$ for each band from the relative phase discussion.
  (e) (f) Relative phases of the bands $\sigma = 41$ and $48$, respectively.
  }
\end{figure}
Next, we explain the calculations for $\ce{S10}$ shown in Fig.~\ref{fig:C10}.
The atomic positions in the unit cell are shown in Figs.~\ref{fig:C10} (a)(b) and this material has approximate $10_7$ screw symmetry, with two helical chains (A and B) and $20$ atoms in one unit cell.
Thus, phonon spectra of $\ce{S10}$ have 60 bands, as shown in Fig.~\ref{fig:C10} (c).
In this paper, we focus on the bands with  indices $\sigma = 41, \cdots, 60$ shown in Fig.~\ref{fig:C10} (d).
{As seen in the band structure, every mode is nearly doubly degenerate, reflecting the two chains in the unit cell. }
Figures~\ref{fig:C10} (e) and (f) show the calculations of the relative phases for the $\sigma=41$ and the $\sigma = 48$ bands, respectively.
The results for the other bands are shown in Appendix~\ref{app:other band}.
{In these calculations we use the wavefunctions in the chain A.}

{This material has been synthesized before, while the calculated phonon spectra have an imaginary part, suggesting that the material is unstable. In fact, our phonon calculation has the following two limitations, by which the imaginary part is unavoidable. 
First, the experimental structure is a high-pressure and high-temperature one, while DFPT calculation can only obtain the phonon spectra for atmospheric pressure and zero temperature. 
Second, the $\ce{S10}$ crystal structure used in our calculation is just 1/4 of the experimentally reported one, 
which have 80 atoms per unit cell. Since each experimental unit cell have 4 left-hand and 4 right-hand helical sulfur chains, with each chain having 10 sulfur atoms, the calculation on the phonon spectra will be very heavy. Here, the crystal structure is approximately described by a smaller unit cell containing only 1 left-handed chain and 1 right-handed chain in our calculation, which has the same symmetry and the same pseudoangular momentum as the experimental structure. Thus we took the smaller unit cell, which does not affect our analysis on the pseudoangular momentum.  
}

{By comparing with the case of exact symmetry, we can assign the value of the quantum number $m$ for the bands $\sigma=41,\ 42, \cdots,\ 54$ as shown in Fig.~\ref{fig:C10} (d), while the bands with $\sigma=55,\cdots, 60$, the quantum number $m$ is ill-defined.}
The value of the pseudoangular momentum $m'$ (associated with $10_7$ symmetry) is calculated by $m' \equiv 7 m \pmod{10}$, 
{which follows from Eq.~(\ref{eq:mml}) for the case with mutually coprime $n$ and $l$.}
These figures show that the quantum number is $m=5$ for the $\sigma=40$ band and changes from $m=-3$ to $m=3$ at $\Gamma$ point for the $\sigma=47$ band.
Similarly, we can assign the values of $m$ for each band as presented in Appendix \ref{app:other band} without ambiguity, and 
the result is shown in Fig.~\ref{fig:C10} (d).
In agreement with our discussion in Sec.~\ref{sec:3D helix}, the quantum number $m \pmod{10}$ increases by $+1$ across the Brillouin zone boundary.
Thus, even when the $10_7$ screw symmetry is broken, the eigenvectors have properties that well reflect the approximate screw symmetry.

\section{\label{sec:conclusion} Conclusion and discussion}

In this paper, we studied how to extract the pseudoangular momentum from the phonon eigenfunction with approximate screw symmetry.
In preparation for this, we considered
a one-dimensional system with partially broken translational symmetry and show 
how to assign the quantum number to each band, which characterizes the relative phase between neighboring atoms. 
We showed that the behavior of this quantum number is naturally understood in the extended Brillouin zone, and showed several key properties of this quantum number and the relative phase. We performed model calculations for a simple case, and extended this method to systems with approximate screw symmetry to extract information about the pseudoangular momentum.
By applying the relative phase defined in this way to the material $\ce{Li6B5}$ with approximate translational symmetry and materials $\ce{SnIP}$ and $\ce{S10}$ with approximate screw symmetry, we confirmed that the information of the quantum number and the pseudoangular momentum {can be
extracted by exact screw symmetries}.
We showed that the assigned pseudoangular momentum is naturally understood in the extended scheme.

{Through our theory and its applications to real materials in this paper, we discuss when the screw symmetry becomes a good symmetry and the pseudoangular momentum becomes well-defined. One can say that the screw symmetry is a
good symmetry, when the energy (or frequency) scale of the symmetry-breaking perturbation is much smaller than the energy scale of the original band structure under exact screw symmetry, which can be a gap size near the band gap, or be a bandwidth. Meanwhile, the calculations in this paper show that it is not so simple whether the screw symmetry is a good symmetry, and the condition for the approximate screw symmetry depends even on the wavenumber and the band considered. It is because 
depending on the band and the wavenumber, the effect of the symmetry-breaking term works differently, as we discussed in Sec.~IV. Whether or not the screw symmetry becomes a good symmetry and the pseudoangular momentum becomes well-defined
is not a simple question in general, and that is why the method to calculate the pseudoangular momentum in this paper is relevant for understanding phonon physics in systems with approximate screw symmetry.
From the analysis in this paper, hybridization between bands with different values of the quantum number $m$ (or the pseudoangular momentum) breaks the screw symmetry and makes $m$ ill-defined, especially at the band crossings. The screw symmetry is largely broken when the hybridization occurs at a lower degree in the Umklapp scattering, and it may depend on the wavevector. 
}

As shown in this study, approximate symmetries can exist in crystals and they affect eigenmodes of the crystals, in addition to exact symmetries indicated by space groups.
For example, in the case of screw symmetry, while only $2$, $3$, $4$, and $6$-fold screw symmetries are allowed as exact symmetries in crystals, we show that approximate $7$-fold and $10$-fold screw symmetries can be realized, which are also reflected in the phonon wave function. {
Since such approximate symmetries are not included in the studies on physical properties, the physics of such approximate symmetries is an interesting topic for further study.
For example, such an approximate symmetry is expected to affect physical processes involving various particles/quasiparticles, such as electronic processes involving multi-phonons/photons and exciton scattering processes. In systems with exact symmetry, symmetry restricts such processes as selection rules, and even when the symmetry is not exact but approximate, selection rules remain valid. While exact screw rotation symmetries are limited in crystals due to the restriction of translational symmetry to twofold, threefold, fourfold and sixfold, an approximate screw symmetry leads to a more variety of multifold screw symmetries such as 7- and 10-fold ones, and they will lead to selection rules that is absent for exact screw symmetries, 
 as has been studied in the context of chiral phonons under exact symmetries, Furthermore, anisotropy in transport properties involving chiral phonons will also reflect the approximate screw symmetry. Angular dependence of phonon transport will lead to $n$-fold anisotropic behavior where $n$ is different from 2, 3, 4 or 6.} 

\begin{acknowledgments}

This work was partly supported by JSPS KAKENHI Grants No.~JP20H04633, JP22H00108 and No.~JP21K13865.

\end{acknowledgments}

\appendix

\section{\label{app:not coprime} Screw symmetry $n_l$ where $n$ and $l$ are not coprime}

In this appendix, we discuss definitions of the quantum numbers $m$ and $m'$ and their relationship under $n_l$ screw symmetry $\hat{C}_{n}^{l}$, when $n$ and $l$ are not coprime. Here $m$ represents a relative phase between neighboring atoms, while $m'$ is associated with an eigenvalue of the $\hat{C}_n^l$ screw operation.
We consider the system with screw symmetry $\hat{C}_{n}^{l}$ with respect to the $z$ axis and translational symmetry  $\hat{T}_{a}$ in the $z$ direction, where $a$ denotes the lattice constant along $z$.

We here consider the case with $n$ and $l$ not being coprime. Let $g(\geq 2)$ denote the greatest common divisor of $n$ and $l$, and we write
\begin{align}
	&n=gn',\ \  l=gl', 
\end{align}
where $n'$ and $l'$ are coprime integers. 
Then the system automatically has 
 $g$-fold rotation symmetry $\hat{C}_{g}$ ($= (\hat{C}_{n}^{l})^{n'} (\hat{T}_{a})^{-l'}$). The translation part of the screw operation $\hat{C}_{n}^{l}$ is
 by $al/n=al'/n'$, which means that the atoms are placed with a period of $a/n'$ along the $z$ direction. Namely, in systems with $n_l$ symmetry, $g$ atoms related by $\hat{C}_g$ symmetry
 share the same $z$ coordinate, and these groups of $g$ atoms are placed along the $z$ axis with a spacing $a/n'$. Thus the number of atoms within the unit cell related by symmetry is $gn'=n$, as expected.

 In the main text, we study the case with $n$ and $l$ being coprime, and we show how to relate two quantum numbers $m$ and $m'$. In this Appendix we 
 estabish the similar relationship between them, when $n$ and $l$ are not coprime.
To this end, we show that there exists a symmetry operation 
 $\hat{\mathcal{O}}\equiv (\hat{C}_n^l)^{p} (\hat{T}_{a})^{q}$ ($p$, $q$: integer) which relates between neighboring atoms, mutually displaced by $\frac{a}{n'}$
  along the $z$ direction. Therefore, integers $p$ and $q$ satisfy $p \frac{la}{n} + q a = \frac{a}{n'}$, i.e. $l' p + n' q = 1$. Since $l'$ and $n'$ are coprime, thanks to the B\'{e}zout's lemma, this equation always has integer solutions for $p$ and $q$.
 Hence, $\hat{\mathcal{O}}=(\hat{C}_n^l)^{p} (\hat{T}_{a})^{q}$ is an operator that rotates the system by $\frac{2 p \pi}{n}$ and translates it by $a/n'$. 
 
 We here note one important point; there are various choices for integers $p$ and $q$, 
 and not all the choices are appropriate for our purpose. In the cases with coprime $n$ and $l$, discussed in the main text, there exists one-to-one correspondence between $m$ and $m'$. In contrast, if they are not coprime, we need to choose $p$ and $q$ properly to guarantee their one-to-one correspondence. For this purpose, we impose a condition that $(\hat{\mathcal{O}})^{r}$ be equal to $\hat{C}_n^l$ modulo translation symmetry, i.e. 
 \begin{align}
 	(\hat{\mathcal{O}})^{r}(\hat{T}_a)^s=\hat{C}_n^l\ \ (r,s:\ \mbox{integer}),
 \end{align}
 which guarantees that the operators $\{\hat{\mathcal{O}},\hat{T}_a\}$ are generators of the screw symmetry $n_l$. 
 This means
 \begin{align}
 	\frac{r}{n'}+s=\frac{l}{n},\ pr\equiv 1 \pmod{n}.
 	\label{eq:alphan}
 \end{align}
The second equation leads to 
$pr+nt=1$ for integers $r$ and $t$. The condition for this equation to have solutions for $r$ and $t$ is that $p$ and $n$ are coprime. As noted earlier there are various choices for $p$, and not all the possible values of $p$ is coprime with $n$. Meanwhile, we can show that among the 
various possible values of $p$, there always exist a value of $p$ that is coprime with $n$. 
To show this, we note that $p$ and $q$ are solutions of $l' p + n' q = 1$, which has solutions of the form $(p,q)=(p_0+n'\kappa,\ q_0-l'\kappa)$ ($\kappa$: integer). Here we note that $p_0$ and $n'$ are coprime because $l' p_0 + n' q_0 = 1$. Thus the set of possible solutions for $p$ forms an arithmetic series, $p=p_0+n'\kappa$ ($p_0$, $n'$: coprime integers), and it contains prime numbers from the Dirichlet's theorem on arithmetic progressions. Thus by choosing $p$ to be a prime number, $p$ and $n$ are coprime. We can also show that the first equation of (\ref{eq:alphan}) is satisfied then.
 
With such a choice of the integers $p$ and $q$, our theory is similar to the case with coprime $n$ and $l$ in the main text. The pseudoangular momentum $m'$ is defined as $\hat{C}_n^l \psi(\vec{r}) = e^{i(kla+2m'\pi)/n} \psi(\vec{r})$ for the wave function $\psi(\vec{r})$ of this system,where 
\begin{align}
	m' \equiv \left\{\begin{array}{ll}
		0, \pm 1, \cdots, \pm (n-1)/2 & \text{:} \ n \ \text{odd} \\
		0, \pm 1, \cdots, \pm (n-2)/2, n/2 & \text{:} \ n \ \text{even}
	\end{array}\right. \pmod n .
\end{align} Then we have
 \begin{align}
 	\hat{\mathcal{O}}\psi(\vec{r})&= (\hat{C}_n^l)^{p} (\hat{T}_{a})^{q} \psi(\vec{r}) \notag \\
 	&= \left(e^{i (lka + 2m'\pi)/n}\right)^{p} \left(e^{i k a}\right)^{q} \psi(\vec{r}) 
 	\label{eq:Opsi2p}
 \end{align}
 holds. Meanwhile, 
 we put the eigenvalue of $\hat{\mathcal{O}}$ to be equal to $ e^{i (gka + 2m\pi)/n}$ because $(\hat{\mathcal{O}})^n=(\hat{T}_a)^g=e^{igka}$, where $m$ 
 is an integer representing the relative phase between neighboring atoms displaced by $a/n'$ along the $z$ axis. Then, from (\ref{eq:Opsi2p}), $m$ is 
 given by $m \equiv m' p \pmod{n}$.
 By using $ pr\equiv 1 \pmod{n}$, we get $m' \equiv mr \pmod{n}$.
 Therefore, we have successfuly established a one-to-one correspondence between the integers $m$ and $m'$.
 In this case, the operators $\{\hat{\mathcal{O}},\hat{T}_a\}$ are generators of the screw symmetry,
and in particular,
 $g$-fold rotation symmetry $C_g$ is expressed in terms of $\hat{\mathcal{O}}$ as $(\hat{\mathcal{O}})^{r n'}(\hat{T}_a)^{-r}=\hat{C}_g$. We summarize properties of these operators in Table \ref{tab:operators}.

\begin{figure}[htb]
	\includegraphics[clip,width=5cm]{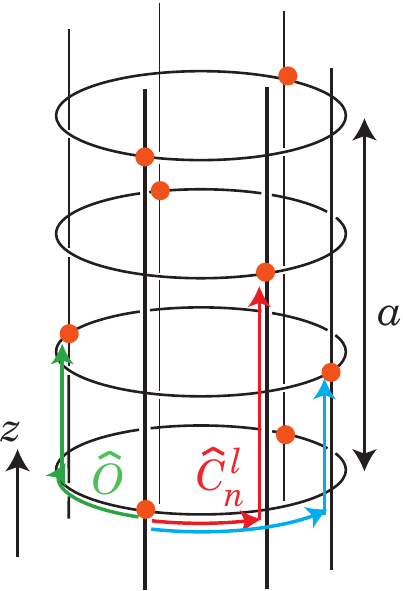}
	\caption{\label{fig:notcoprime}
	Schematic figure for the case with $6_4$ screw symmetry, which corresponds to 
	$n=6$ and $l=4$ being non-coprime. The orange dots represent positions of atoms, and the red arrow represents the screw operation $\hat{C}_6^4$. The green arrow represents the operation $\hat{\mathcal{O}}$, which connects between atoms in the neighboring layers. It seems that there are two choices for the operator $\hat{\mathcal{O}}$ that relates neighboring atoms displaced by $a/3$ along the $z$ axis, but we cannot choose the blue arrow as the operator $\hat{\mathcal{O}}$, because it does not generate all the screw operations; in other words, the operation by the blue arrow does not connect all the atoms in the unit cell.
	}
\end{figure}
We show an example of $n=6$, $l=4$ in Fig.~\ref{fig:notcoprime}.  Here we have $n'=3$ and $l'=2$, and two solutions of
$l'p+n'q=1$ are $(p,q)=(-1,1),(2,-1)$, and the operator $\hat{\mathcal{O}}$ defined by these choices of $(p,q)$ are shown by the green arrow and the blue arrow, respectively. In our discussion above, we need to take $p$ to be coprime with $n$, so only the former choice of $(p,q)=(-1,1)$ (green arrow) is allowed. By this proper choice of $\hat{\mathcal{O}}$, all the atoms are related by $\hat{\mathcal{O}}$. On the other hand, if we take the other choice of $(p,q)=(2,-1)$, not all the atoms are related by $\hat{\mathcal{O}}$. Thus the eigenvalue of $\hat{\mathcal{O}}$ does not possess the same infomation as that of $\hat{C}_n^l$. 

With such a proper choice of $\hat{\mathcal{O}}$, we name the atoms in the unit cell like $(i,j)$ ($i,j$: integer) in the following way. 
Within the unit cell (e.g. Fig.~\ref{fig:notcoprime}), we begin with the $g$ atoms with the smallest $z$ coordinate, and we name them 
$(1,1)$, $(1,2)$, $\cdots, (1,g)$, in the way that the $(1,j)$ atom is transformed to $(1,j+1)$ by the $g$-fold rotation $\hat{C}_g$. Then 
we name other atoms so that the operator $\hat{\mathcal{O}}$ transforms the $(i, j)$ atom to $(i+1, j)$.
Then, we express the eigenvectors $\bm{\epsilon}$ of the dynamical matrix in the form $\bm{\epsilon} = (\vec{\epsilon}_{1, 1}, \cdots, \vec{\epsilon}_{1, g}, \cdots, \vec{\epsilon}_{n', 1}, \cdots, \vec{\epsilon}_{n', g})^{\mathrm{T}}$, where $\vec{\epsilon}_{i, j}$ is the displacement of the atom $(i, j)$ $(1\leq i\leq n',1\leq j \leq g)$.
We define te relative phase for this eigenvector $\bm{\epsilon}$ and explain how to use it to extract the pseudoangular momentum $m'$.

In this case, for the eigenvector $\bm{\epsilon}^{(\sigma)}$ for the $\sigma$-th band, we define the relative phase $\theta_{i, j, \alpha}$ as 
\begin{align}
  z_{i, j, \alpha}^{(\sigma)} (k) &= \left\{ \begin{array}{ll}
    u_{i+1, j, \alpha}^{(\sigma)} / u_{i, j, \alpha}^{(\sigma)} & (i = 1, \cdots n'-1) \\
    u_{1, j+p, \alpha}'^{(\sigma)} e^{ika} / u_{n', j, \alpha}^{(\sigma)} & (i = n')
  \end{array} \right. , \\
  \theta_{i, j, \alpha}^{(\sigma)}(k) &= \arg z_{i, j, \alpha}^{(\sigma)} (k) ,
\end{align}
where 
$\alpha = x, y, z$,
$\vec{u}_{i, j}^{(\sigma)} = A^{i-1}_p \vec{\epsilon}_{i, j}$,
$\vec{u}_{1, j, \alpha}'^{(\sigma)} = A^{n'}_p  \vec{\epsilon}_{1, j}$ and
$A_p$ is a $2p\pi/n$ rotation matrix (\ref{eq:Ap}).
This relative phase is given by $\theta_i(k)=(gka + 2 m \pi)/n$ when the system has exact $n_l$ screw symmetry.
Then, even in the case where the $n_l$ screw symmetry is slightly broken, we can extract the information of the quantum number $m$ and the pseudoangular momentum $m'$
from the eigenvector  $\bm{\epsilon}^{(\sigma)}$.

\begin{table}
	\begin{tabular}{llll}
		Operator & Rotation & Translation & \\
				 &\ \ \   angle\ \ \  &  &  \\ \hline
		$\hat{C}_{n}^{l}$ & $\frac{2\pi}{n}$&$\frac{l'}{n'}a$ & $=(\hat{\mathcal{O}})^r(\hat{T})^s$\\
			$\hat{\mathcal{O}}$ & $\frac{2\pi p}{n}$&$\frac{1}{n'}a$ & $=(\hat{C}_n^l)^p(\hat{T})^q$\\
			$\hat{C}_g$ & $\frac{2\pi }{g}$&$0$ & $=(\hat{\mathcal{O}})^{rn'}(\hat{T})^{-r}=(\hat{C}_n^l)^{n'}(\hat{T})^l$\\ \hline
		\end{tabular}
	\caption{Relationships between various operations in the screw symmetry. The operators 
		$\{\hat{C}_{n}^{l},\hat{T}_a\}$ generate the screw symmetry $n_l$, and so do the operators
	$\{\hat{\mathcal{O}},\hat{T}_a\}$
}
	\label{tab:operators}
\end{table}

\section{\label{app:other band} Relative phase of the other bands for two materials, $\ce{SnIP}$ and $\ce{S10}$ }

In this appendix, we show the results of calculations for the relative phases of $\ce{SnIP}$ and $\ce{S10}$ for bands not shown in the main text.

\begin{figure}[h]
  \includegraphics[clip,width=\linewidth]{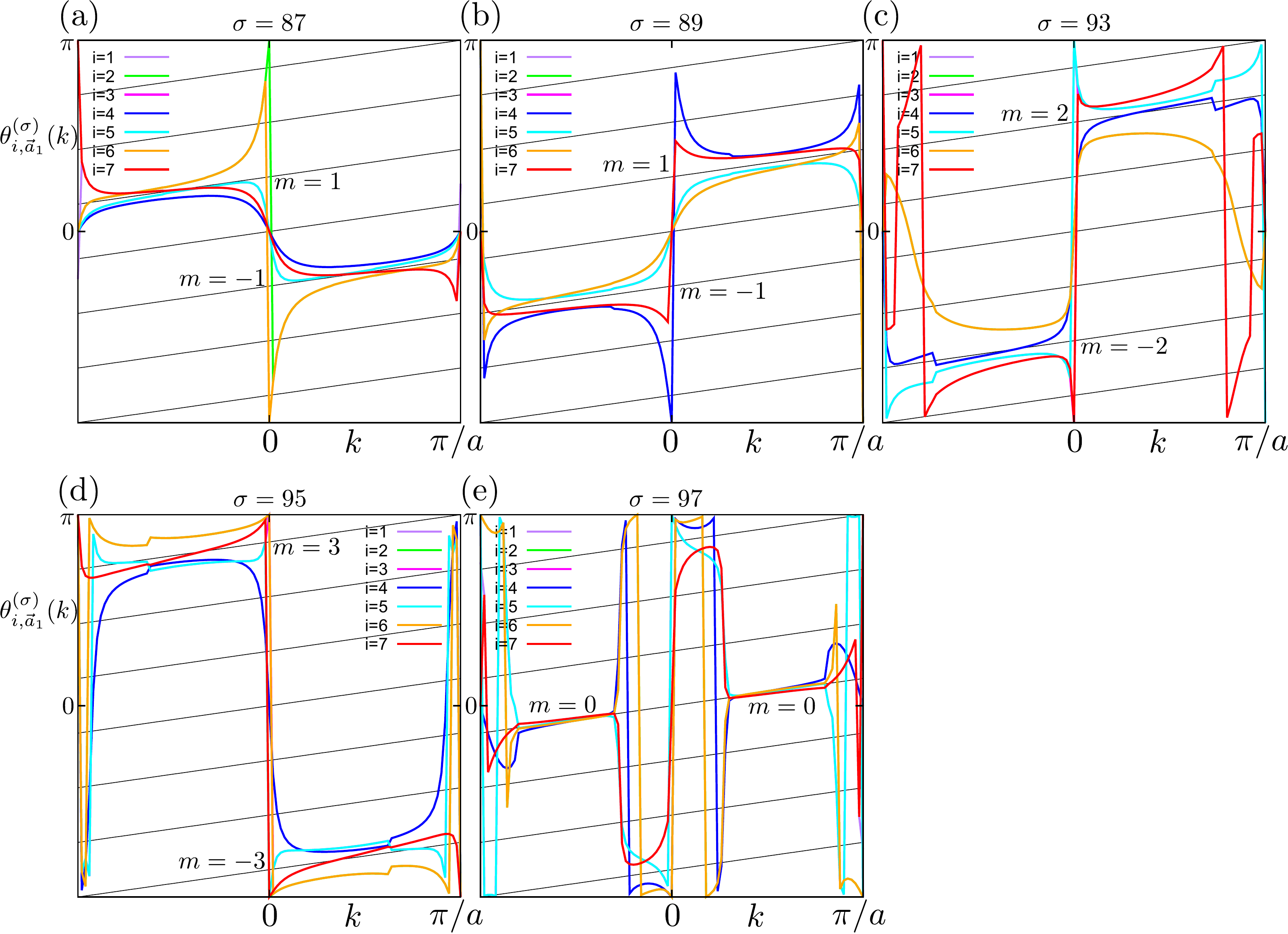}
  \caption{\label{fig:app SnIP}
  Relative phase of $\ce{SnIP}$.
  (a), (b), (c), (d) and (e) correspond to the relative phases of the bands $\sigma=87$, $89$, $93$, $95$ and $97$ in Fig.~\ref{fig:SnIP} (d), respectively.
  }
\end{figure}

First, we show the relative phase of $\ce{SnIP}$ in Fig.~\ref{fig:app SnIP}.
Figures~\ref{fig:app SnIP} (a), (b), (c), (d) and (e) correspond to the relative phases of the bands ($\sigma=87$, $89$, $93$, $95$ and $97$) in Fig.~\ref{fig:SnIP} (d) in the main text, respectively.
From this result, we can extract the quantum number $m$, summarized in Fig.~\ref{fig:SnIP}.
{We note that the relative phase for the $\sigma=97$ band in Fig.~\ref{fig:app SnIP}(e) shows a crossover of the value of $m$ as $0\rightarrow  -3\rightarrow 3\rightarrow 0$.
It is due to hybridization between the bands with $m=\pm 3$ and that with $m=0$ at their anticrossings, as seen around 8THz in
Fig.~\ref{fig:SnIP} (c).  Thus at the anticrossing between bands, the properties of the wavefunction of the bands are exchanged. }

\begin{figure}[htb]
	\includegraphics[clip,width=\linewidth]{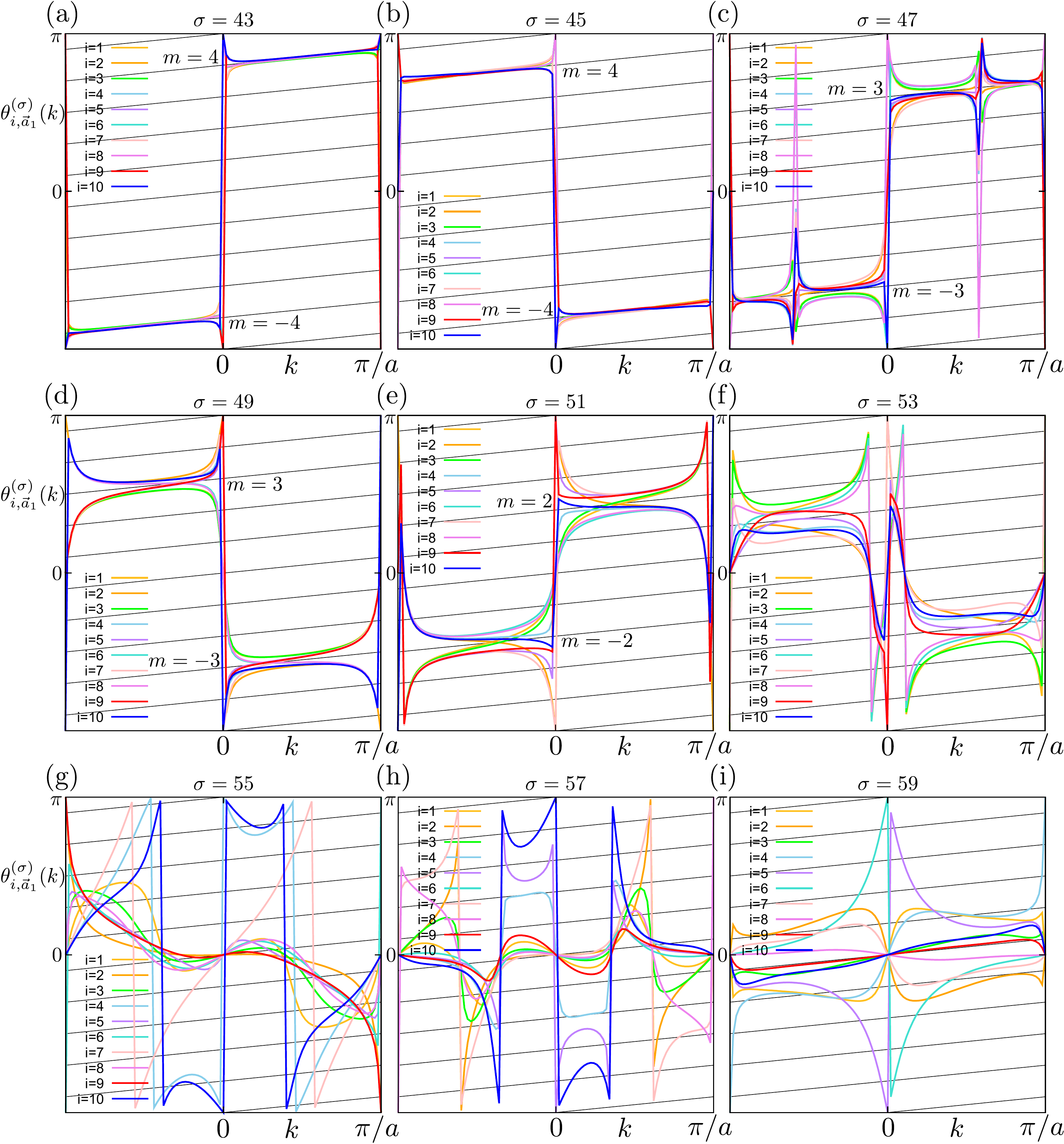}
	\caption{\label{fig:app C10}
		Relative phase of $\ce{S10}$.
		(a), (b), (c), (d), (e), (f), (g), (h) and (i) correspond to the relative phases of the bands $\sigma=43$, $45$, $47$, $49$, $51$, $53$, $55$, $57$ and $59$ in Fig.~\ref{fig:C10} (d), respectively.
	}
\end{figure}

Finally, we show the relative phase of $\ce{S10}$ in Fig.~\ref{fig:app C10}.
Figures~\ref{fig:app C10} (a), (b), (c), (d), (e), (f), (g), (h) and (i) correspond to the relative phases of the bands ($\sigma = 43$, $45$, $47$, $49$, $51$, $53$, $55$, $57$ and $59$) in Fig.~\ref{fig:C10} (d) in the main text, respectively.
From this result, we can extract the quantum number $m$ {for $\sigma=41,\ 42, \cdots,\ 54$, summarized in Fig.~\ref{fig:C10}, while for the bands with $\sigma=55,\cdots, 60$, the quantum number $m$ is ill-defined.}

\begin{figure}[htb]
	\includegraphics[clip,width=\linewidth]{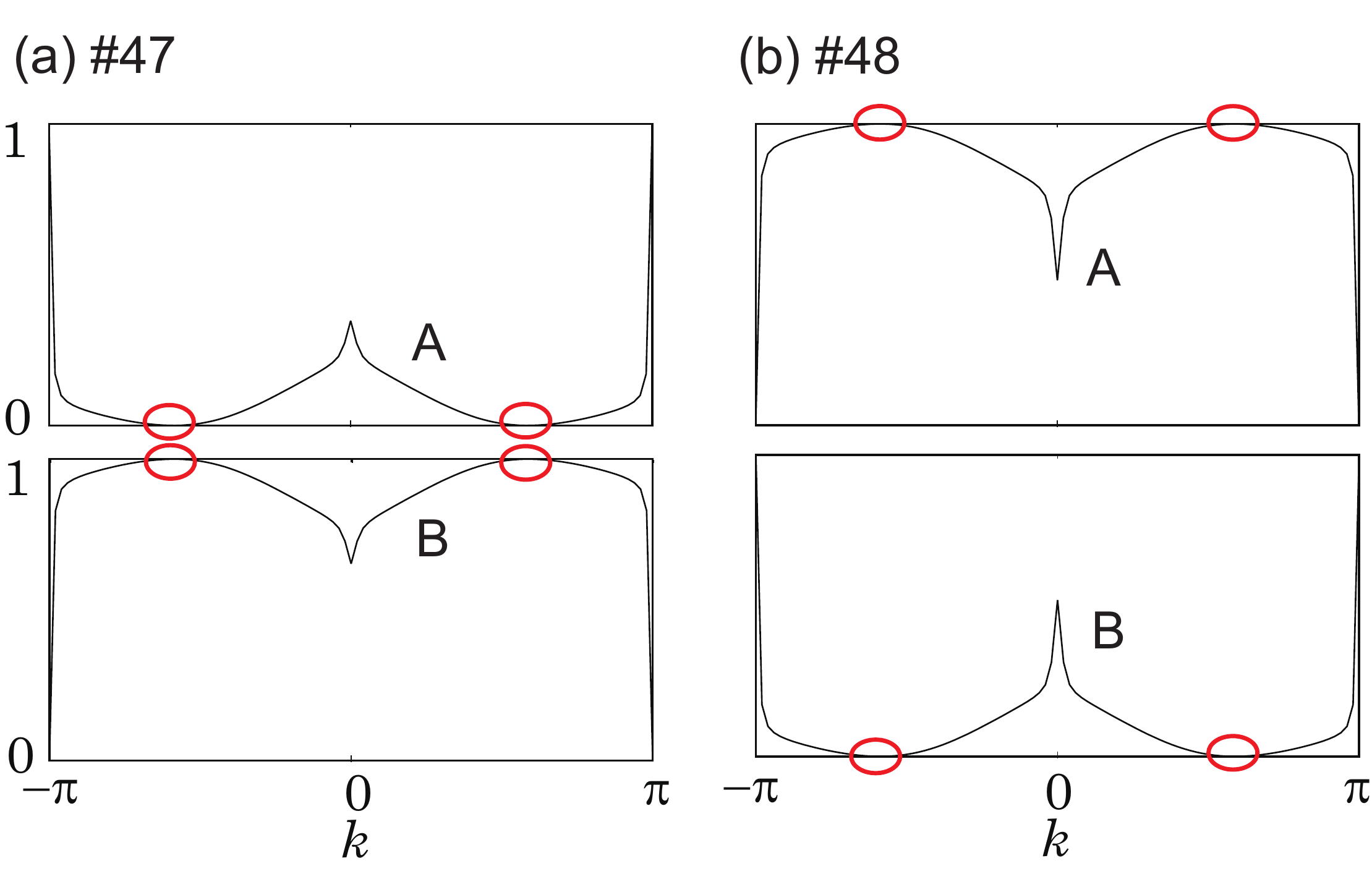}
	\caption{\label{fig:S10norm}
		Norm of the phonon modes within the chains A and B for the phonon modes with (a) $\sigma=47$ and (b) $\sigma=48$ in the material $\ce{S10}$. At the wavenumvers $k\sim \pm 0.6\pi$ (red circles), the modes are localized only in one chain.	}
\end{figure}

{By comparing the results for the two bands $\sigma=48$ (Fig.~\ref{fig:C10}(f)) and $\sigma=47$ (Fig.~\ref{fig:app C10} (c)), which are almost degenerate in the band structure, 
we notice that the latter has an anomalous behavior near $k=\pm 0.6\pi$, while the former does not. One might think that this anomaly comes from anticrossing between bands, but it is not the case, since such an anomaly from anticrossing should appear simultaneously in the two bands 
$\sigma=47,48$. Instead, this anomaly comes from an anomalous behavior of hybridization between phonon modes in the two chains A and B, as we explain in the following. In fact, at the wavenumber $k=\pm 0.6\pi$, the $\sigma=47$ band is localized only in the chain B, while the $\sigma=48$ band only in the chain A. It is seen in 
Fig.~\ref{fig:S10norm} where the sum of the squares of the phonon amplitudes from the normalized eigenvalues within each chain is plotted for the two bands. 
It is noted that in our calculation of the quantum number $m$ for S$_{10}$, we used the wavefunctions in the chain A. Then in the phonon mode $\sigma=47$ in Fig.~\ref{fig:app C10} (c), the phonon amplitudes in the chain A is almost zero, which gives rise to the anomaly around $k=\pm 0.6\pi$, because at this point the result becomes highly sensitive to details of the system and to numerical errors. We also checked that if we use the phonon amplitude in the chain B instead, an anomaly appears in the $\sigma=48$ band. Thus, this anomaly in the relative phases is attributed to the localization of the eigenmodes into a single chain.}

Here, one may wonder why this localization of eigenmodes into a single chain happens in this case. 
In general, even when two chains exist within the unit cell, the phonon modes within the two chains are hybridized and they are never localized within 
a single chain. In the present case, we find that it is allowed from the $C_{2x}T$ symmetry, where $C_{2x}$ is the twofold rotation symmetry with respect to the axis {along $\vec{a}_2$ perpendicular to the chains (see Fig.~\ref{fig:C10}(a)(b))}, and $T$ is the time-reversal symmetry. {In the present material with two chains, A and B, this $C_{2x}T$ symmetry is preserved within each chain, and} the {$N\times N$} phonon dynamical matrix $D_i(k)$ satisfies
\begin{align}
	MD_i(k)M^{\dagger}=D_i(k)^{*},
\end{align}
where {$N(=10)$ is the number of atoms within the unit cell in one chain, $i=A,B$ indicates the chains A and B}, $k$ is the wavenumber along the chain, and $M$ is a {$N\times N$} matrix representing the $C_{2x}$ operation. For example, in the present case, for the phonons with atomic displacements along the $z$ axis, $M$ is given by
\begin{align}
&	M=\begin{pmatrix}
		&&&&-1\\
		&&&-1&\\
		&&\cdots &&\\
        &-1&&&\\
		-1&&&&
		\end{pmatrix}.
\end{align}
Under this symmetry, when $u_i(k)$ is an eigenvector at the wavenumber $k$, the vector $(Mu_i(k))^{*}$ is also an eigenvector having the same eigenfrequency, and therefore they are proportional to each other: $u_i(k)\propto (Mu_i(k))^{*}$. By properly choosing the phase of $u_i(k)$, one can always make $u_i(k)$ to satisfy $u_i(k)=(Mu_i(k))^{*}$. Thereby, the dynamical matrix $D_i(k)$
is diagonalized by a {$N\times N$} unitary matrix $U_i(k)$, $U_i(k)^{\dagger}D_i(k)U_i(k)=\varepsilon_i(k)$, satisfying 
$U_i(k)=(MU_i(k))^{*}$, where $\varepsilon_i(k)$ is a diagonal matrix with the eigenfrequencies as diagonal elements. Next, 
{if we combine the two chains without hybridization}, the dynamical matrix for the entire system
\begin{align}
	&D(k)=\begin{pmatrix}
		D_A(k)&\\&D_B(k)
	\end{pmatrix}
\end{align}
is diagonalized by a {$2N\times 2N$} unitary matrix $U(k)$ satisfying
\begin{align}
	&U(k)=\begin{pmatrix}
		U_A(k)&\\&U_B(k)
	\end{pmatrix},\\
&U_i(k)^{\dagger}D_i(k)U_i(k)=\varepsilon_i(k), \\ 
&U_i(k)=(MU_i(k))^{*}\ \ (i=A,B),
\end{align}
Next, we introduce an interchain coupling preserving $C_{2x}T$ symmetry. Then the dynamical matrix becomes
\begin{align}
	&D'(k)=\begin{pmatrix}
		D_A(k)&V(k)\\ V(k)^{\dagger}&D_B(k)
	\end{pmatrix},
\end{align}
where {$V(k)$ is a $N\times N$ matrix} satisfying $MV(k)M^{\dagger}=V(k)^{*}$. Then, after unitary transformation by $U(k)$, the dynamical matrix becomes
\begin{align}
	&	U(k)^{\dagger}D'(k)U(k)=\begin{pmatrix}
		\varepsilon_A(k)&V'(k)\\V'(k)^{\dagger}&\varepsilon_B(k)
	\end{pmatrix}\equiv d(k),
\end{align}
where $V'(k)=U_A(k)^{\dagger}V(k)U_B(k)$. 
Then from the $C_{2x}T$ symmetry, we get $V'(k)=V'(k)^{*}$. Therefore the matrix $d(k)$ is a
real matrix with its diagonal elements representing the eigenfrequencies in each chain, and the off-block-diagonal elements in $V'(k)$ represent a hybridization between the modes in the chains A and B. In the present case the spectra of the chains A and B are the same, and we focus on the particular eigenmodes with the 
frequency $\omega_n$. Then we can approximate the dynamical matrix by retaining only the matrix elements involving these modes with the frequency $\omega_n$, and the reduced dynamical matrix is
\begin{align}
	&	\bar{d}(k)=\begin{pmatrix}
		\omega_n(k)&v'(k)\\v'(k)&	\omega_n(k)
	\end{pmatrix},
\end{align} where $v'(k)$ is a real parameter, representing the interchain hybridization. Only when $v'(k)=0$, the two chains are decoupled. Because $v'(k)$ is real, if $v'(k)$ has 
a different sign between $k=0$ and $k=\pi$, it necessarily goes across zero, where the two chains are decoupled and the phonon modes are localized only in a single chain, either A or B. It explains the anomalous behavior shown in Fig.~\ref{fig:S10norm}. 
We note that without this $C_{2x}T$ symmetry, the hybridization $v'(k)$ becomes complex in general, and it cannot be zero when changing only one parameter $k$. Thus, to summarize, under the $C_{2x}T$ symmetry, it can happen at a certain value of $k$
that an interchain hybridization becomes zero and the eigenmodes are localized only in a single chain. 
Among the bands considered, the same behavior is seen also in the bands $\sigma=41,42$ around $k=\pm 0.8\pi$ and in
the bands $\sigma=45,46$ around $k=0.6\pi$, but not in other bands, and this is consistent with our scenario.  

\end{document}